\documentclass{aa}  

\usepackage{graphicx}
    \graphicspath{{figures/}}
\usepackage{txfonts}
\usepackage{lipsum}
\usepackage{subcaption}         % necessary for continued figures, example in section 3
\usepackage{lscape}             % to rotate a single page table, example in appendix.
\usepackage{placeins}           % useful with \FloatBarrier, to keep 
\usepackage{csquotes}
\usepackage{subcaption}
\usepackage{ulem}

\usepackage[colorlinks=true,linkcolor=blue,citecolor=blue,urlcolor=blue]{hyperref}
\newcommand{\RXGru}{RX~Gru}
\newcommand{\Solaris}{Solaris}
\newcommand{\TESS}{\textit{TESS}}
\newcommand{\Kepler}{\textit{Kepler}}
\newcommand{\SWASP}{SWASP}
\newcommand{\Gaia}{\textit{Gaia}}
\newcommand{\MESA}{\textsc{mesa}}
\newcommand{\creme}{CR\'{E}ME}
\newcommand{\lightcurve}{\textsc{lightkurve}}
\newcommand{\ie}{i.e.\ }
\newcommand{\eg}{e.g.\ }

\begin{document}

   \title{\RXGru{}: a short-period pre-main-sequence eclipsing binary with a distant circumbinary companion}

   %\subtitle{Subtitle}

%%%%%%%%%%%%%%%%%%%%%%%%%%%%%%%%%%%%%%%%
% Please do not include ORCIDs next to author names.
% Only ORCIDs authenticated by individual authors in EDP Sciences editorial system will be taken into account.
% ORCIDs included here will be removed.
%%%%%%%%%%%%%%%%%%%%%%%%%%%%%%%%%%%%%%%%

   \author{F.\ Marcadon\inst{1}
        \and A.\ Moharana\inst{2}%\fnmsep\thanks{Shows the usage of elements in the author field}
        \and T.\ B.\ Pawar\inst{3}
        \and G.\ Pawar\inst{4}
        \and K.\ G.\ He{\l}miniak\inst{4}
        \and J. P. Marques\inst{5}
        \and M.\ Konacki\inst{4}
        }

   \institute{Nicolaus Copernicus Astronomical Center, Polish Academy of Sciences, ul.\ Bartycka 18, 00-716 Warszawa, Poland\\
             \email{fmarcadon@camk.edu.pl}
             %\thanks{Shows the usage of elements in the author field}
            \and Astrophysics group, Keele University, ST5 5BG Staffordshire, UK
            \and Department of Astrophysics and Planetary Sciences, Villanova University, 800 East Lancaster Avenue, Villanova, PA 19085, USA
            \and Nicolaus Copernicus Astronomical Center, Polish Academy of Sciences, ul.\ Rabia\'{n}ska 8, 87-100 Toru\'{n}, Poland
            \and Institut d'Astrophysique Spatiale, UMR8617, CNRS, Universit\'{e} Paris-Saclay, B\^{a}timent 121, 91405 Orsay Cedex, France\\ }

   %\date{Received September 30, 20XX}
   \date{}

% \abstract{}{}{}{}{}
% 5 {} token are mandatory
 
%  \abstract
%   {Optional, leave empty if necessary.  The heading “Context” is used when needed to
%give background information on the research conducted in the paper}
%   {Mandatory. The objectives of the paper are defined here.} 
%   {Mandatory. The methods of the investigation are outlined here}
%   {Mandatory. The results are summarized here.}
%   {Optional, leave empty if necessary.  “Conclusions” can be used to
%explicit the general conclusions that can be drawn from the paper.}

\abstract{We report the discovery of a new short-period pre-main-sequence eclipsing binary, \RXGru{}, orbited by a distant circumbinary companion. We characterized the system by analysing the photometric observations from the \Solaris{} network, the \textit{Transiting Exoplanet Survey Satellite}, and the Super Wide Angle Search for Planets survey, combined with the radial velocities from four high-resolution spectrographs: HARPS, FEROS, CHIRON, and HRS. We derived the parameters of the eclipsing components, which are $M_{\rm Aa} = 1.004^{+0.027}_{-0.026}\,$M$_\odot$, $R_{\rm Aa} = 1.007\pm0.021\,$R$_\odot$, and $T_{\rm eff,Aa} = 5379\pm289\,$K for the primary, and $M_{\rm Ab} = 0.985^{+0.024}_{-0.025}\,$M$_\odot$, $R_{\rm Ab} = 1.024\pm0.023\,$R$_\odot$, and $T_{\rm eff,Ab} = 5322\pm278\,$K for the secondary. We determined the age of the system from the observed parameters using two evolution codes, \MESA{} and Cesam2k20. We obtained an age of $\sim$28$\,$Myr, placing the two stars at the very end of the pre-main-sequence phase. We also derived the minimum mass and orbital period of the tertiary companion, which are found to be $M_{\rm B} = 89.0\pm3.5\,$M$_{\rm Jup}$ and $P_{\rm AB} = 23.79 ^{+0.10 }_{-0.25}\,$yr, respectively. We conclude that \RXGru{} consists of a tight inner binary composed of two twin components and an outer low-mass companion (a massive brown dwarf or a very low-mass star) in a relatively wide orbit, and we suggest that the system was formed via the dynamical unfolding mechanism coupled with the shared accretion of the circumbinary material by the binary components.}


   \maketitle

%\tbf{BOLD: New text}

%\boldm{$BOLD\!\!: New\,text\,in\,math\,mode$}

%{\red RED: Comments}

%%%%%%%%%%%%%%%%%%%%%%%%%%%%%%%%%%%%%%%%%%%%%%%%%%

%%%%%%%%%%%%%%%%% INTRODUCTION %%%%%%%%%%%%%%%%%%%

\section{Introduction}

It is now well established that a large fraction of solar-type stars form in binaries or higher-order multiple systems, namely $\sim$35 and 10 per cent of the total population, respectively \citep{2010ApJS..190....1R}. However, the dominant mechanism that leads to the formation of the shortest period binary systems, \ie those with $P \lesssim 20\,$d, is still a subject of debate. Many studies argue that close binaries form predominantly in hierarchical triples through Kozai--Lidov \citep{1962AJ.....67..591K,1962P&SS....9..719L} cycles and tidal friction (hereafter KCTF; see \eg \citealt{2016ComAC...3....6T}, for a review). There are, however, several limitations of the KCTF mechanism in reproducing certain observed features of the binary population. First of all, the KCTF mechanism occurs only at initial relative inclinations between 39.2$^\circ$ and 140.8$^\circ$. Recent surveys by \citet{2024A&A...685A..43M} and \citet{2025A&A...695A.209B} studied the distribution of mutual inclinations in hierarchical triple systems observed by \Kepler{} and the \textit{Transiting Exoplanet Survey Satellite} (\TESS{}). They found that most of their triple systems have a nearly coplanar configuration, implying that the KCTF mechanism cannot be the dominant source of short-period binaries. Secondly, the KCTF mechanism acts on particularly long time-scales ($\gtrsim$100$\,$Myr; \citealt{2007ApJ...669.1298F,2014ApJ...793..137N}) that are incompatible with the formation of young (<30$\,$Myr) short-period pre-main-sequence (PMS) binaries (see \citealt{2014NewAR..60....1S,2020ApJ...902..107L}, and references therein). Finally, the KCTF mechanism alone cannot reproduce the observed excess of short-period binaries with component mass ratios close to unity (hereafter referred as twins; \citealt{2000A&A...360..997T}). Indeed, the origin of such twins might result from the shared accretion of the circumbinary material by the binary components during the PMS phase \citep{2018ApJ...854...44M}, while the triple system is still being stabilized by the dynamical unfolding mechanism proposed by \citet{2012Natur.492..221R}.

In this context, the detection of an increasing number of young, well-characterized hierarchical triples is an essential step toward our understanding of short-period binary formation. To this end, eclipsing binaries (EBs) that are also double-lined spectroscopic binaries provide a direct determination of the stellar parameters through their dynamics. Indeed, it is possible to measure the masses and radii of each component of a double-lined EB with an exquisite precision, better than $\sim$1--3 per cent, and to derive its age without resorting to more advanced stellar modelling \citep{2021MNRAS.508.5687H}. Additionally, the so-called eclipse timing variation (ETV) method has already proven its potential for identifying hierarchical triple stellar systems \citep{2022Galax..10....9B} and substellar circumbinary companions \citep{2016A&A...587A..82W,2018A&A...620A..72W,2021A&A...647A..65W} from both space-based and ground-based surveys. Unfortunately, only a few short-period PMS EBs are known to belong to hierarchical triples (see fig.~7 of \citealt{2020ApJ...902..107L}). This lack of PMS EBs was also noted by \citet{2024A&A...690A.153M} from a sample of 48 compact ($P_{\rm out} \lesssim 1000\,$d) hierarchical triples. In this paper, we announce the discovery of a new PMS EB, namely \RXGru{}, which consists of two nearly equal-mass components in a short-period orbit of $P=0.743\,$d (General Catalogue of Variable Stars version 5.1; \citealt{2017ARep...61...80S}). The main photometric and astrometric properties of \RXGru{} (\Gaia{} DR3 6542838829517539328) are given in Table~\ref{tab:properties}. We also detected a circumbinary companion (a massive brown dwarf or a very low-mass star) orbiting the eclipsing pair with a period of $\sim$23.8$\,$yr. This makes \RXGru{} a potentially interesting target for studying the early formation processes of short-period binaries in hierarchical triple systems.

This article is organized as follows: Section~\ref{sec:obs} describes the observational data used in this work, including \Solaris{}, \TESS{}, and Super Wide Angle Search for Planets (SuperWASP) photometry, as well as radial-velocity (RV) measurements of \RXGru{}. Section~\ref{sec:analysis} presents the spectral and orbital analysis of the system leading to the determination of the fundamental properties (mass, radius, effective temperature) of the two eclipsing components and to the detection of the circumbinary companion. In Section~\ref{sec:discussion}, we discuss the implications of the main features of \RXGru{} on its evolutionary status. Finally, the conclusions of this work are summarized in Section~\ref{sec:summary}.

\begin{table*}
\centering
    \caption{Main photometric and astrometric properties of \RXGru{}.}
    \label{tab:properties}
    \renewcommand{\arraystretch}{1.0}
    \begin{tabular}{@{}lccc@{}}
        \hline\hline
        Parameter                                       & Value                             & Reference                   \\
        \hline
        RA J2000                                        & 22:58:16.71                       & --                          \\
        Dec J2000                                       & $-$41:49:34.24                    & --                          \\
        $G$ (mag)                                       & 10.432                            & \citet{2022yCat.1355....0G} \\
        $G_{\rm BP}$ (mag)                              & 10.828                            & \citet{2022yCat.1355....0G} \\
        $G_{\rm RP}$ (mag)                              & 9.853                             & \citet{2022yCat.1355....0G} \\
        $G-V$ (mag)\tablefootmark{a}                    & $-0.205$                          & --                          \\
        $V$ (mag)\tablefootmark{a}                      & 10.637                            & --                          \\
        $V_{\rm ASAS}$ (mag)\tablefootmark{b}           & 10.680                            & \citet{2012AcA....62...67K} \\
        $V_{\rm syst}$ (mag)\tablefootmark{c}           & $10.658\pm0.030$                  & --                          \\
        $T_{\rm eff,\,Ab}/T_{\rm eff,\,Aa}$             & 0.980\tablefootmark{d}            & --                          \\
        $E(B - V)$ (mag)                                & 0.05                              & \citet{2018yCat.2354....0G} \\
        $A_{\rm V}$ (mag)                               & 0.20                              & \citet{2018yCat.2354....0G} \\
        $A_{\rm V}$ (mag)                               & 0.25\tablefootmark{e}             & --                          \\
        $\pi$ (mas)                                     & $6.935\pm0.017$                   & \citet{2022yCat.1355....0G} \\
        RUWE\tablefootmark{f}                           & 0.899                             & \citet{2022yCat.1355....0G} \\
        \texttt{astr\_ex\_noise} (mas)\tablefootmark{g} & 0.074                             & \citet{2022yCat.1355....0G} \\
        \texttt{vis\_periods\_used}\tablefootmark{g}    & 17                                & \citet{2022yCat.1355....0G} \\
        \hline
    \end{tabular}
    \tablefoot{
    \tablefoottext{a}{The $G-V$ colour and $V$ magnitude were calculated using \Gaia{} DR3 photometry and the relations from \citet{2021A&A...649A...3R}.\\}
    \tablefoottext{b}{$V$ magnitude from the All Sky Automated Survey (ASAS; \citealt{2002AcA....52..397P}).\\}
    \tablefoottext{c}{The adopted value of $V_{\rm syst}$ was computed as the mean of the values derived from \Gaia{} DR3 and ASAS photometry.\\}
    \tablefoottext{d}{From the light-curve modelling presented in Section~\ref{sec:LC_analysis}.\\}
    \tablefoottext{e}{From the spectral energy distribution (SED) fitting presented in Section~\ref{sec:teff}.\\}
    \tablefoottext{f}{RUWE stands for renormalized unit weight error \citep{2021A&A...649A...2L}.\\}
    \tablefoottext{g}{Abbreviations for \texttt{astrometric\_excess\_noise} and \texttt{visibility\_periods\_used} \citep{2021A&A...649A...2L}.}
    }
\end{table*}

%%%%%%%%%%%%%%%%%%%%%%%%%%%%%%%%%%%%%%%%%%%%%%%%%%

%%%%%%%%%%%%%%%%% OBSERVATIONS %%%%%%%%%%%%%%%%%%%

\section{Observations}
\label{sec:obs}

\subsection{Solaris photometry}
\Solaris{} (PI: M.\ Konacki) is a global network of telescopes consisting of four fully autonomous observatories located in the Republic of South Africa (Solaris-1 and -2), Australia (Solaris-3), and Argentina (Solaris-4). Each observatory consists of a telescope with a primary mirror of 0.5-m diameter focusing on CCDs with a resolution of $2048 \times 2048$ pixels, thermoelectrically cooled to $-70^\circ$~Celsius. The filter wheels allow for multi-colour photometry in Johnson and Sloan bands. The network commissioning, its hardware, software and processing capabilities are described in \cite{Kozlowski_2017}.

RX~Gru was a program target in the first run of the telescope, collecting observations of the star from 2015 to 2017. After a preliminary analysis, it was revisited in 2021 to take the total span of \Solaris{} observations to 6$\,$yr.  The minimum cadence in most of the observed eclipses is 1$\,$min, with older photometry having cadences up to 3$\,$min. 

The data were reduced with the dedicated \Solaris{} pipeline which calculates fluxes using a pseudo point spread function photometric method \citep{2024MNRAS.527...53M}. The pipeline extracts photometry from astrometry-corrected frames to provide the final light curve with barycentric time corrections based on \cite{eastman2014} and was implemented using \textsc{barycorrpy}\footnote{\url{https://pypi.org/project/barycorrpy/}}.

\subsection{\TESS{} photometry}
Photometry from \TESS{} \citep{2015JATIS...1a4003R} was used to obtain the light curves. The mission scans almost the entire sky, subdivided into sectors, each observed for about 27$\,$d on average. For \RXGru{} (TIC~152825521), we used the 2-min and 10-min cadence photometry from sector~1 (2018 July 25--August 22)\footnote{Through Guest Investigator programmes G011083 and G011154.} and sector~28 (2020 July 31--August 25), respectively.
%\RXGru{} has a \TESS{} magnitude of 9.92$\,$mag. 
The different versions of data are available in the Barbara A.\ Mikulski Archive for Space Telescopes (MAST)\footnote{\url{https://mast.stsci.edu/}}, namely Simple Aperture Photometry (SAP) and Pre-search Data Conditioning SAP (PDCSAP) fluxes. 

We used SAP fluxes for sector~1 in our analysis, which are fetched using the \lightcurve{} package \citep{2018ascl.soft12013L}. To extract the data from sector 28, we used the \lightcurve{} package along with the TESSCut tool \citep{tesscut2019ascl.soft05007B} to access the \TESS{} full-frame image (FFI) cutouts. We applied customised apertures to measure the fluxes and corrected these fluxes by subtracting the background scattered light captured by the detectors. Then, we normalized the light curves by dividing by the median flux value of each sector.
    
\subsection{SuperWASP photometry}

For our analysis, we also used the photometric observations from the SuperWASP survey (hereafter \SWASP{}; \citealt{2006PASP..118.1407P,2010A&A...520L..10B}). The \SWASP{} survey was a dedicated search for transiting exoplanetary systems using two wide-field robotic telescopes, one located at the Observatorio del Roque de los Muchachos on La Palma, the other at the Sutherland Station of the South African Astronomical Observatory (SAAO). Observations of \RXGru{} (1SWASP~J225815.57$-$414926.0) were performed from 2006 May 7 to November 12 and from 2007 May 7 to November 13 through broad-band filters of 400--700$\,$nm. The typical cadence of the observations was 9--12$\,$min.

The light curve was downloaded from the \SWASP{} public archive\footnote{\url{https://www.superwasp.org/}} and consists of 11\,107 flux measurements with their corresponding errors. We identified a total of 41 well-covered eclipses (19 primary and 22 secondary) in the light curve. In the following, we will take advantage of these data to extend the time span of the ETV observations for the system (see Section~\ref{sec:rv_etv}).

\subsection{High-resolution spectroscopy} \label{sec:spectro}

The \RXGru{} system was observed with four high-resolution spectrographs: HARPS \citep[$R\sim120\,000$;][]{harps_mayor} mounted behind the ESO-3.6m telescope in La silla (Chile), FEROS \citep[$R\sim44\,000$;][]{feros_kaufer} at the MPG-2.2m telescope in La Silla, CHIRON \citep[$R\sim28\,000$, fiber mode;][]{chiron_schwab,chiron_tokovinin} at the SMARTS-1.5m telescope in CTIO (Chile), and HRS \citep[$R\sim65\,000$;][]{crause_hrs} fed by the SALT telescope in SAAO (South Africa). 

The first observations were done with HARPS and FEROS in 2004 (under engineering runs) and 2006, and are available in the ESO Archive. We used HARPS to observe the target in 2011 and 2012 as part of the Comprehensive Research with \'{E}chelles on the Most interesting Eclipsing binaries \citep[\creme;][]{creme_pasm}. These observations constitute the bulk of our spectroscopic data. %More recently, we re-observed the target with CHIRON \tbf{(April 2022, June 2023) and HRS (August-October 2022).}
More recently, we re-observed the target with CHIRON (April 2022) and HRS (August-September 2022).
All the data were reduced with dedicated pipelines, provided by the observatories. 
%Excluding the observations taken during or near the eclipses, we ended up with \tbf{19 spectra (HARPS: 10, CHIRON: 4, HRS: 3, FEROS: 2)} used for RV measurements, and 10 (HARPS only) used for disentangling.
Excluding the observations taken during or near the eclipses, we ended up with 15 spectra (HARPS: 10, CHIRON: 1, HRS: 2, FEROS: 2) used for RV measurements, and 10 (HARPS only) used for disentangling.

%The RVs were measured with our own implementation of the TODCOR routine \citep{todcor}, with synthetic template spectra, and individual measurement errors calculated with a bootstrap procedure \citep{helm12}. The RVs are collected in the Table~\ref{tab:RV_obs} in the Appendix.
The RVs were measured with our own implementation of the TODCOR routine \citep{todcor}. We used echelle order with wavelengths longer than $\sim$4000\,\AA, except for CHIRON ($>\,$4500\,\AA), which had a very weak signal in the blue part. As templates, we used synthetic spectra computed with ATLAS~9 \citep{kurucz}, which do not reach wavelengths longer than 6500$\,$\AA. This lowers the number of useful echelle orders, but reduces the influence of telluric lines, and cuts off the broad H$\alpha$ line. We also did not consider orders with the sodium D lines ($\sim$5900\,\AA), as they are often affected by the interstellar medium. We used templates created for $T_{\rm eff} = 5200\,$K, $\log g = 4.5\,$dex, solar metallicity, and rotationally broadened to $v \sin i =30$\,km\,s$^{-1}$. The individual measurement errors were calculated with a bootstrap procedure \citep{helm12}, which is sensitive to the signal-to-noise ratio (S/N) of a component, and velocity of rotation. The rotational broadening of lines and their profile variability due to spots were the main sources of measurement errors. The RVs are collected in the Table~\ref{tab:RV_obs} in the Appendix.

%%%%%%%%%%%%%%%%%%%%%%%%%%%%%%%%%%%%%%%%%%%%%%%%%%

%%%%%%%%%%%%%%%%% ANALYSIS %%%%%%%%%%%%%%%%%%%%%%%

\section{Analysis}
\label{sec:analysis}

\subsection{Determination of times of minima}
\label{sec:ecl_times}

%As part of our analysis, we also 
We determined the mid-eclipse times of \RXGru{} by applying the timing procedure described in \citet{2024ApJ...976..242M} to the \TESS{}, \Solaris{}, and \SWASP{} light curves of the system. We summarize here the main steps of the procedure, and we refer the reader to \citet{2024MNRAS.527...53M} and \citet{2025arXiv250220626P} for recent applications.

We adopted the phenomenological model of \citet{2015A&A...584A...8M}, which analytically describes the eclipse profile as
\begin{equation}
    f(t_i,\Theta) = \alpha_0 + \sum^{n_{\rm e}}_{k=1} \alpha_k \, \psi(t_i,T_k,d_k,\Gamma_k,C_k),\label{eq:ecl_profile}
\end{equation}
where $\alpha_0$ is the flux zero-point shift, $n_{\rm e}$ is the number of eclipses during one cycle, and $\alpha_k$ is a scaling coefficient of the eclipse profile function, which is written as
\begin{multline}
    \psi(t_i,T,d,\Gamma,C) = \Bigg\{ 1 + C \bigg( \frac{t_i-T}{d} \bigg)^2 \Bigg\} \\ \times \Bigg\{ 1-\bigg\{ 1-\exp \bigg[ 1-\cosh \bigg( \frac{t_i-T}{d} \bigg) \bigg] \bigg\}^\Gamma \Bigg\}.
\end{multline}
Here, $T$, $d$, $\Gamma$, and $C$ are the time of minimum, the eclipse width, the kurtosis, and the scaling parameter, respectively. For each individual eclipse, the time of minimum is estimated as
\begin{equation}\label{eq:ecl_mod}
    \begin{split}
        T_k & = T_{0,k} + P E + \Delta_k \\
        & = T_{0,k} + P \times {\rm round} \bigg( \frac{t_i-T_{0,k}}{P} \bigg) + \Delta_k,
    \end{split}
\end{equation}
where $T_{0,k}$, $P$, $E$, and $\Delta_k$ are the reference time of eclipse, the orbital period, the epoch, and the observed minus calculated ($O-C$) time difference, respectively. As discussed in detail later in Section~\ref{sec:spots}, the light curve of the system exhibits a rapidly varying O’Connell effect, which refers to the height difference between successive out-of-eclipse maxima \citep{1951PRCO....2...85O,1968AJ.....73..708M}. This asymmetry may introduce significant biases in the measured mid-eclipse times, as shown by \citet{2006Ap&SS.304..363M} and \citet{2024ApJ...976..242M}. Given that the light curve is unstable, we decided to fit each eclipse individually, and we excluded the out-of-eclipse section of the light curve from the fit. In order to account for the apparent asymmetry of the eclipse profiles, we added to our model a linear combination of three cosine functions. 

\begin{figure}
    \includegraphics[trim = 2.8cm 1.95cm 3.0cm 2.0cm,clip,width=1.0\columnwidth,angle=0]{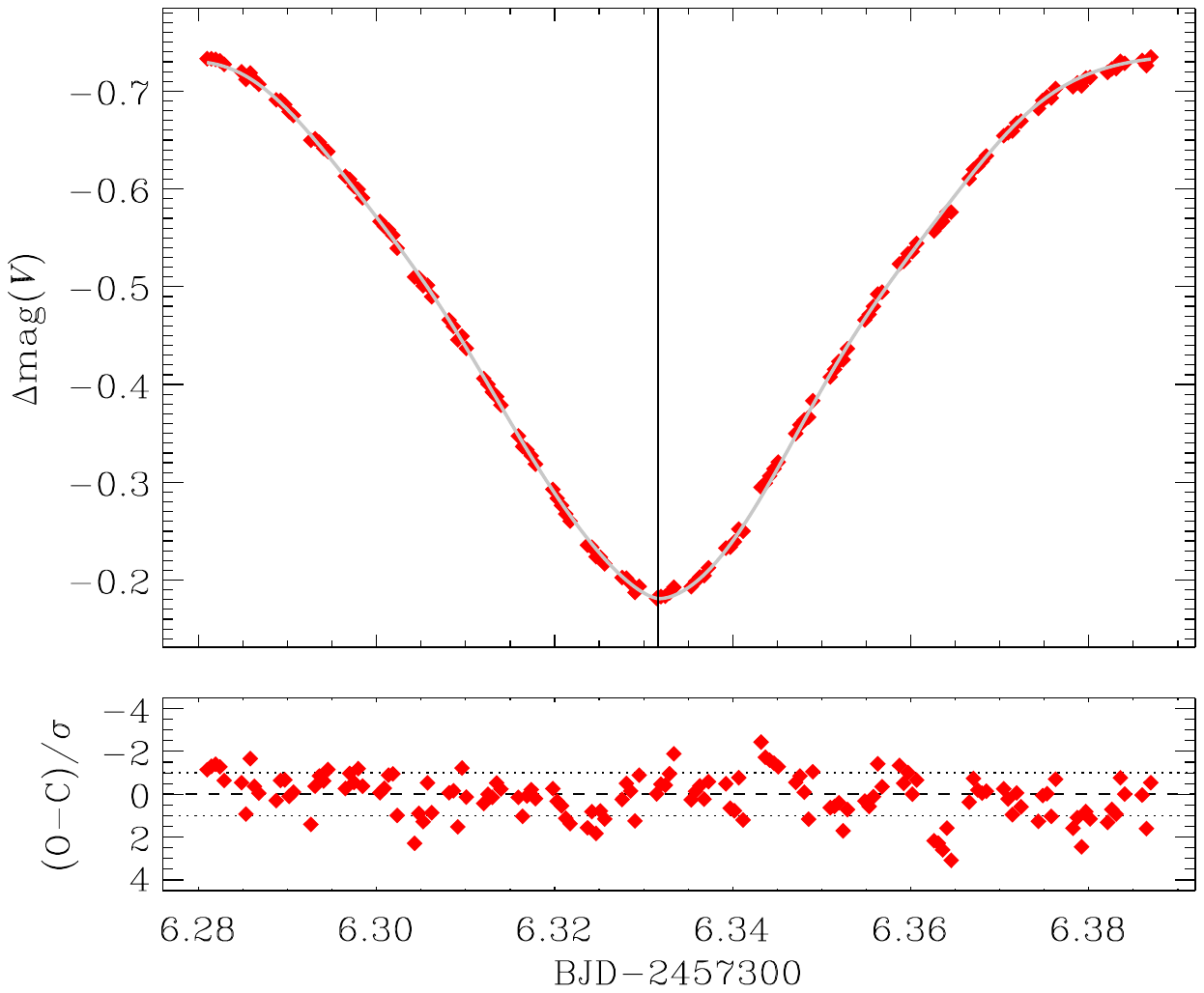}
    \caption{\Solaris{} photometry of the primary eclipse of \RXGru{} observed in the $V$ band on 2015 October 10. Red diamonds denote the observations, while the grey line corresponds to the best-fitting model determined for each eclipse using the procedure described in Section~\ref{sec:ecl_times}. The vertical line indicates the time of minimum light associated with the best-fitting model (see Table~\ref{tab:ecl_times} in Appendix~\ref{app:ecl_times}). Fitting residuals are shown in the lower panel.}
    \label{fig:ecl_times}
\end{figure}

We then performed a Markov Chain Monte Carlo (MCMC) fit of each individual eclipse using the aforementioned model. The best-fitting model of a single primary eclipse is shown in Fig.~\ref{fig:ecl_times}, and the times of minima derived from our fitting procedure are listed in Table~\ref{tab:ecl_times} in Appendix~\ref{app:ecl_times}, along with their 1$\sigma$ errors.

\subsection{Combined RV and ETV curve analysis}
\label{sec:rv_etv}

In this section, we present the methodology employed to process
the RV and ETV measurements of \RXGru{}. Our analysis is based on new and archival RV measurements, which are described in Section~\ref{sec:spectro}, as well as on the times of minima reported in Section~\ref{sec:ecl_times}.

\begin{figure}
    \includegraphics[trim = 2.5cm 1.93cm 3.0cm 2.0cm,clip,width=1.0\columnwidth,angle=0]{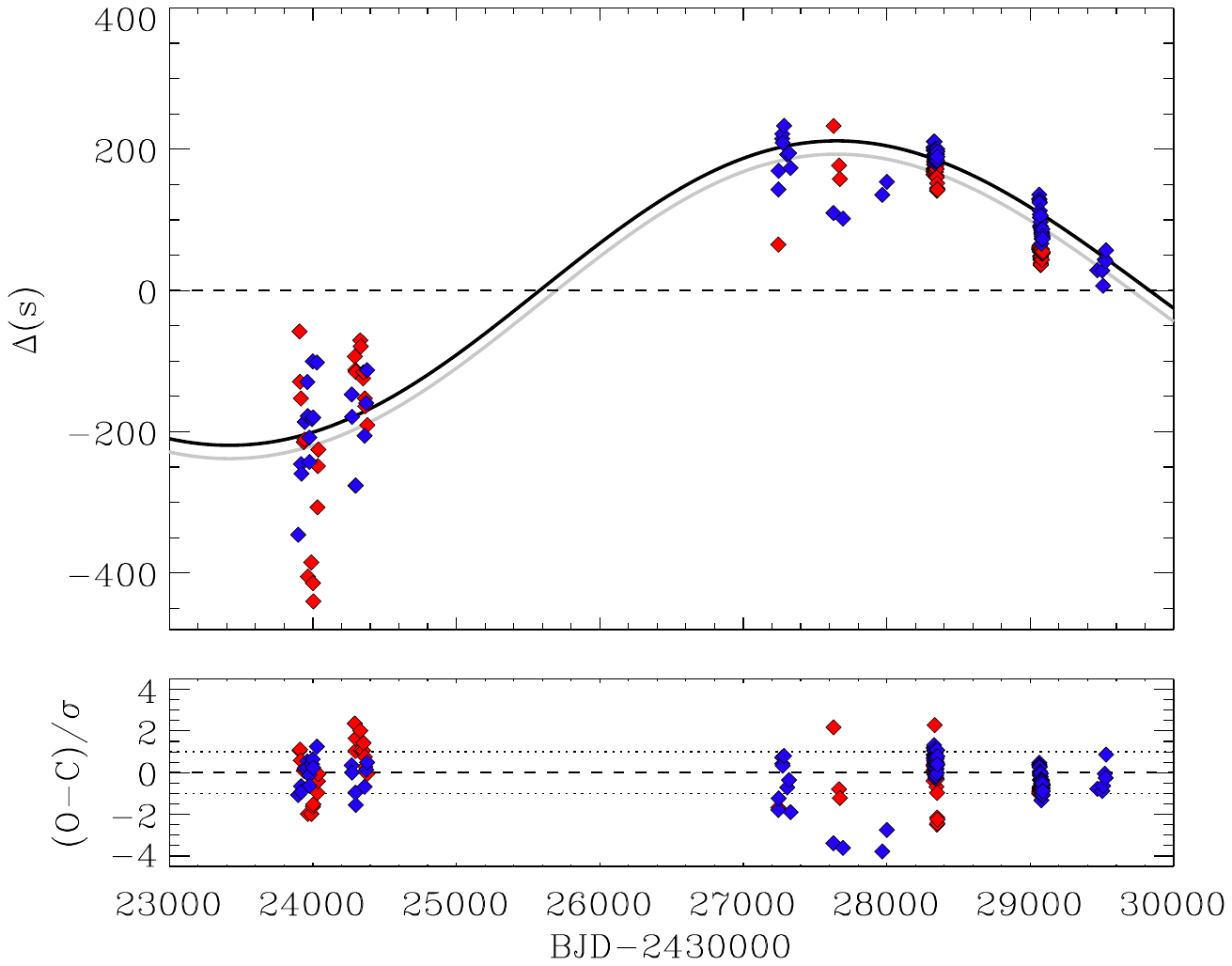}
    \caption{ETVs for \RXGru{}. The primary and secondary eclipse times are indicated by blue and red symbols, respectively, while the primary and secondary ETV solutions are shown as black and grey solid lines, respectively. Fitting residuals are shown in the lower panel.}
    \label{fig:ETV}
\end{figure}

In order to derive the orbital parameters of the system, we adopted the Bayesian approach presented in \citet{2018A&A...617A...2M,2020MNRAS.499.3019M} and \citet{2024ApJ...976..242M}. We defined the global likelihood of the data given the orbital parameters as
\begin{equation}
{\cal L}={\cal L}_{\rm RV} \, {\cal L}_{\rm ETV},
\end{equation}
where ${\cal L}_{\rm RV}$ and ${\cal L}_{\rm ETV}$ are the likelihoods of the RV and ETV data respectively, computed from
\begin{equation}
    \ln {\cal L}_{\rm RV} = -\frac{1}{2}\,\sum^{N_{\rm RV}}_{i=1} \, \left(\frac{V_i^{\rm mod}-V_i^{\rm obs}}{\sigma_{V,i}} \right)^2,
\end{equation}
\begin{equation}
    \ln {\cal L}_{\rm ETV} = -\frac{1}{2}\,\sum^{N_{\rm ETV}}_{i=1} \, \left(\frac{\Delta_i^{\rm mod}-\Delta_i^{\rm obs}}{\sigma_{\Delta,i}} \right)^2.
\end{equation}
$N_{\rm RV}$ and $N_{\rm ETV}$ denote the number of available RV and ETV observations, respectively, and $\sigma$ refers to the associated uncertainties. The terms $V$ and $\Delta$ stand for the RVs and the $O-C$ time differences, respectively. The exponents `mod' and `obs' refer to the modelled and observed constraints used during the fitting procedure. We computed the observed $O-C$ time differences as
\begin{equation}
    \Delta_{\rm obs} = T_{\rm o}(E) - T_{\rm c}(E) = T_{\rm o}(E) - T_0 - P E,
\end{equation}
where $T_{\rm o}(E)$ and $T_{\rm c}(E)$ refer to the observed and calculated times of minima at epoch $E$, respectively. The values of $T_0$ and $P$ are taken from Table~\ref{tab:RV_param}. The resulting $O-C$ diagram is shown in Fig.~\ref{fig:ETV}. Both the primary and secondary minima of \RXGru{} show a periodic variation in the $O-C$ diagram, which may be caused by the presence of an unseen third body in the system. In this case, ETVs are generally attributed to the light-traveltime effect (LTTE; \citealt{1990BAICz..41..231M}), also known as the R\o{}mer delay, or to the effect of dynamical perturbations occurring in the system (see \eg \citealt{2015MNRAS.448..946B}). For \RXGru{}, we expect the dynamical ETV contribution to be negligible with respect to the LTTE contribution due to the large outer-to-inner period ratio of the system (see equation~12 of \citealt{2016MNRAS.455.4136B}).
%${\mathcal A}_{\rm dyn} = 54.2^{+1.8}_{-2.0}\,$ms.
We thus modelled the ETVs in the mathematical form of LTTE \citep{2015MNRAS.448..946B,2016MNRAS.455.4136B}:
\begin{equation}
    \Delta_{\rm mod} = c_0 + c_1 E - \frac{a_{\rm A}\sin i_2}{c} \, \frac{\big(1-e_2^2\big)\sin(\upsilon_2+\omega_2)}{1+e_2\cos \upsilon_2},
    \label{eq:etv_model}
\end{equation}
where $c_0$ and $c_1$ are factors that correct the respective values of $T_0$ and $P$ for the ETV effect, $a_{\rm A}$ is the semimajor axis of the EB's barycentric orbit, $c$ is the speed of light, and $i_2$, $e_2$, $\upsilon_2$, and $\omega_2$ are the inclination, eccentricity, true anomaly, and argument of periastron of the third component's relative orbit, respectively. For simplicity, we use the semi-amplitude of the LTTE ETVs defined as \citep{1952ApJ...116..211I}
\begin{equation}
    A_{\rm LTTE} = \frac{a_{\rm A}\sin i_{\rm AB}}{c} = \frac{{\mathcal A}_{\rm LTTE}}{(1-e_2^2 \cos^2 \omega_2)^{1/2}}.\label{eq:Altte}
\end{equation} 
In our analysis, we also accounted for the possibility of a slightly eccentric inner orbit, which results in a measurable displacement of the secondary ETV curve with respect to the primary one (see Fig.~\ref{fig:ETV}). Indeed, after a first round of fits, we noticed that the secondary ETV residuals were systematically shifted from the primary ones by $\sim$19$\,$s. Thus, for secondary eclipses, the term $-\,(T_{\rm s}-T_{\rm p})+0.5\,(P+c_1)$ has to be added to the right-hand side of equation~(\ref{eq:etv_model}). It corresponds to the time interval between the primary and secondary eclipses (\citealt{1959cbs..book.....K,2001icbs.book.....H}):
\begin{equation}
    \frac{2 \pi \, (T_{\rm s}-T_{\rm p})}{P+c_1} = \pi + 2 \tan^{-1} \frac{e_1 \cos \omega_1}{(1-e_1^2)^{1/2}} + \frac{2 e_1 \cos \omega_1 \, (1-e_1^2)^{1/2}}{(1-e_1^2 \sin^2 \omega_1)},
\end{equation}
where $e_1$ and $\omega_1$ are the eccentricity and the argument of periastron of the eclipsing pair, respectively. Finally, we computed the RVs of the three components using a double-Keplerian model, as described in \citet{2020MNRAS.499.3019M}. This model consists of the parameters $\mathcal{P}_{\rm orb} = (c_0,c_1,K_{\rm Aa},K_{\rm Ab},T_{\rm A},e_{\rm A},\omega_{\rm A},P_{\rm AB},T_{\rm AB},e_{\rm AB},\omega_{\rm AB},\gamma_{\rm AB},A_{\rm LTTE})$, where subscripts Aa and Ab refer, respectively, to the primary and secondary components and subscript AB refers to the relative orbit between the eclipsing pair A and the third body B. We use this notation from now on, which is more adequate to describe a hierarchical triple system. For the derivation of these orbital parameters, we employed an MCMC method using the Metropolis--Hastings algorithm \citep{1953JChPh..21.1087M,10.1093/biomet/57.1.97}. Briefly, the procedure consists of setting 10 chains of 10 million points each with starting points taken randomly from appropriate distributions. The new set of orbital parameters is here computed using a random walk as follows:
\begin{equation}
{\cal P}_{\rm orb}^{t'}={\cal P}_{\rm orb}^{t}+\alpha_{\rm rate} \, \Delta {\cal P}_{\rm orb},
\end{equation}
where $\Delta {\cal P}_{\rm orb}$ is given by a multivariate normal distribution with independent parameters and $\alpha_{\rm rate}$ is an adjustable parameter that is reduced by a factor of 2 until the rate of acceptance of the new set exceeds 25 per cent, marking the end of the burn-in phase (first 10 per cent of each chain). We then derived the posterior probability distribution of each parameter from the Markov chains after rejecting the initial burn-in phase. For all parameters, we computed the median and the credible intervals at 16 per cent and 84 per cent, corresponding to a 1$\sigma$ interval for a normal distribution. 

\begin{figure*}
    \includegraphics[trim = 2.5cm 1.93cm 3.0cm 2.0cm,clip,width=1\columnwidth,angle=0]{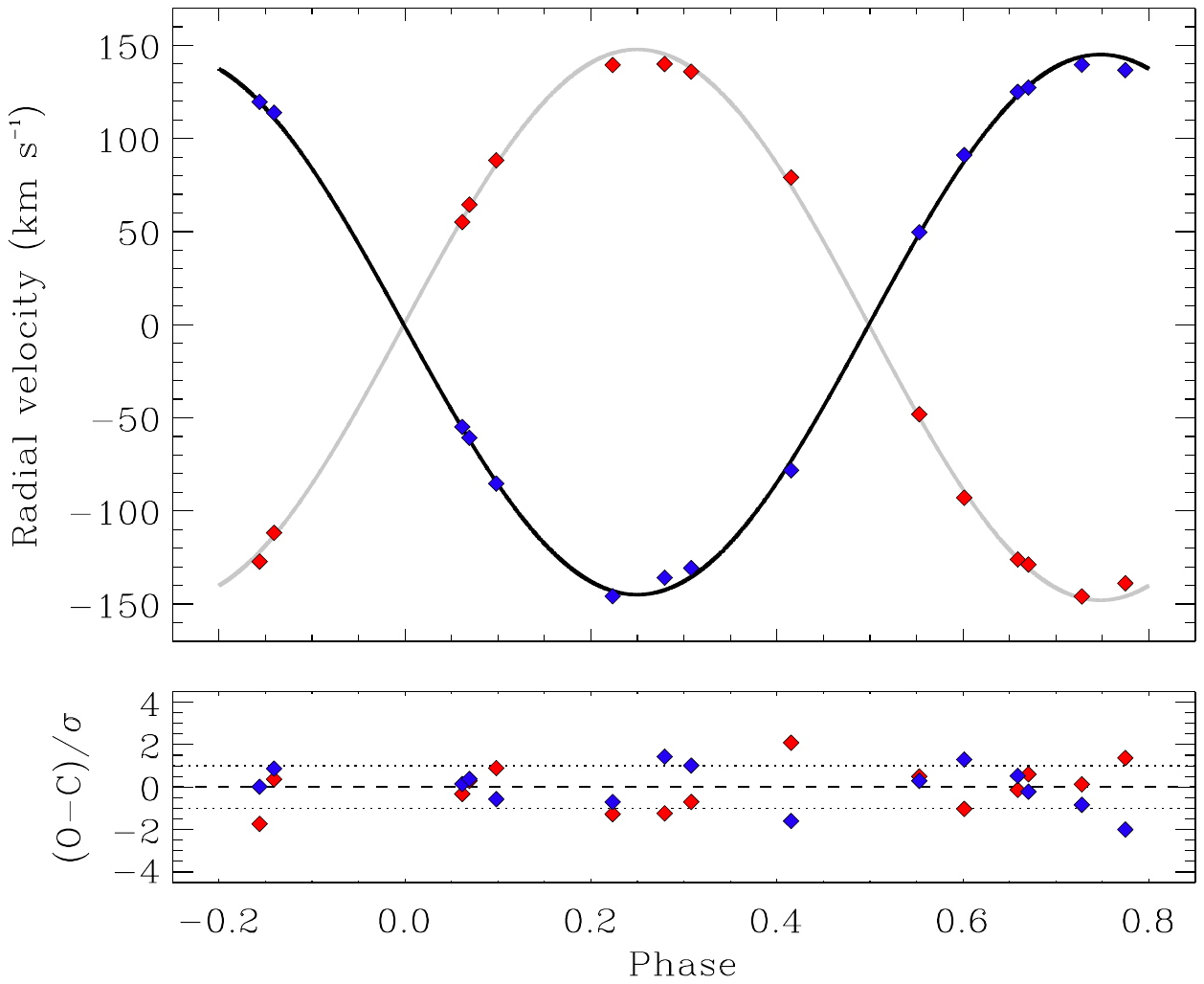}
    \includegraphics[trim = 2.5cm 1.93cm 3.0cm 2.0cm,clip,width=1\columnwidth,angle=0]{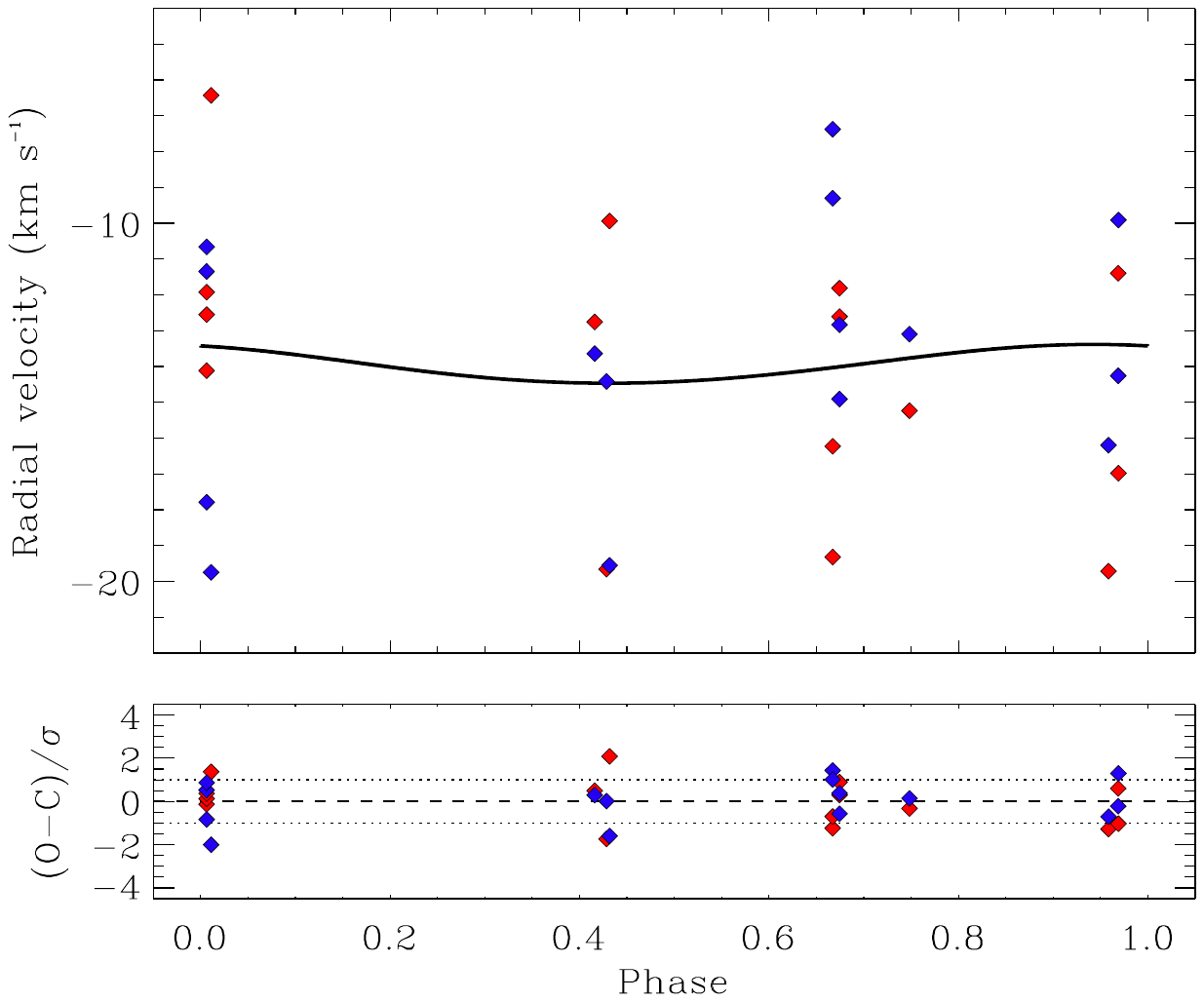}
    \caption{RV curves of \RXGru{} described by a double-Keplerian orbital model using RVs of stars Aa (blue) and Ab (red). Upper-left panel: Best-fitting solutions for stars Aa (black) and Ab (grey) after having removed the 23.8-yr modulation induced by the third body. The curve is phase-folded at the orbital period $P_{\rm A}=0.743\,$d, where phase 0 is set for the time of primary minimum $T_0$. Upper-right panel: Best-fitting solution for the centre of mass of the eclipsing pair after having removed the orbital motion of stars Aa and Ab. The curve is phase-folded at the orbital period $P_{\rm AB}\!\simeq 23.8\,$yr, where phase 0 is set for the time of periastron passage $T_{\rm AB}$. The corresponding RV semi-amplitude is expected to be small compared to the scatter of the RVs (see the text). Lower panels: Residuals of the fitting procedure.}
    \label{fig:RV_phase}
\end{figure*}

An essential aspect when combining different data types is the determination of a proper relative weighting. To this end, we performed a preliminary set of fits and computed the normalized chi-squared ($\chi^2/N$) of the best fit for the primary and secondary ETV curves and for the primary and secondary RV curves. Then, we rescaled the error bars of the eclipse times and RV measurements by multiplying them by the corresponding value of $\sqrt{\chi^2/N}$ and performed a new fit of the ETV and RV curves. We also fitted three additional terms to take into account the zero-point differences between the four spectrographs. Here, we assumed that the shift is the same for the two stars. We obtained FE$-{\rm HA} = 0.09\pm0.31$\,km\,s$^{-1}$, CH$-{\rm HA} = 0.21\pm0.45$\,km\,s$^{-1}$, and HR$-{\rm HA} = -0.27\pm0.31$\,km\,s$^{-1}$. Finally, we checked that the best-fitting solution has $\chi^2/N \sim 1$ for all the curves and that the mean error on the measurements is of the same order of magnitude as the root mean square (rms) of the fitting residuals. For ETV measurements, we found $\sigma_{\rm p} = 39$\,s and $\sigma_{\rm s} = 60$\,s while for RV measurements, we found $\sigma_{\rm Aa} = 3.5$\,km\,s$^{-1}$ and $\sigma_{\rm Ab} = 3.7$\,km\,s$^{-1}$. The best-fitting solution for the whole system is shown in Figs.~\ref{fig:ETV} and~\ref{fig:RV_phase} and the corresponding orbital parameters are given in Table~\ref{tab:RV_param}. We point out that the RV signal caused by the third component is not detected in our data due to the small value of the semi-amplitude $K_{\rm A}$ ($\sim$0.5$\,$km\,s$^{-1}$) compared to the scatter of the RVs ($\sim$3.5--3.7$\,$km\,s$^{-1}$).

%\begin{equation*}
%{\rm Pri.\ Min.} = {\rm BJD}\,2458325.91804 + 0.\!\!^{\rm d}743140442 \cdot E.
%\end{equation*}

\begin{table*}
    \centering
    %\begin{minipage}{85mm}
        \caption{Orbital parameters and derived quantities for the best-fitting model of the RV and ETV data.}
        \label{tab:RV_param}
        {
        \renewcommand{\arraystretch}{1.0}
        \begin{tabular}{@{}lccc@{}}
            \hline\hline
             &  & 84 per cent & 16 per cent \\
            Parameter & Median & interval & interval \\
            \hline
            $K_{\rm Aa}$ (km\,s$^{-1}$) & 145.0 & +1.6 & $-$1.6 \\
            $K_{\rm Ab}$ (km\,s$^{-1}$) & 147.9 & +1.8 & $-$1.7 \\
            $P_{\rm A}$ (d) & 0.743140442\tablefootmark{a} & +0.000000051 & $-$0.000000046 \\
            $T_0$ (BJD$-245\,0000$) & 8325.91804\tablefootmark{a} & +0.00021 & $-$0.00006 \\
            $T_{\rm A}$ (BJD$-245\,0000$) & 8322.591 & +0.030 & $-$0.011 \\
            $e_{\rm A}$ & 0.0031 & +0.0051 & $-$0.0019 \\
            $\omega_{\rm A}$ ($^\circ$) & 278.6 & +14.8 & $-$5.5 \\
            $A_{\rm LTTE}$ (s) & 214.8 & +6.9 & $-$7.7 \\
            $K_{\rm A}$ (km\,s$^{-1}$)\tablefootmark{b} & 0.543 & +0.017 & $-$0.019 \\
            $P_{\rm AB}$ (d) & 8690 & +36 & $-$91 \\
            $P_{\rm AB}$ (yr) & 23.79 & +0.10 & $-$0.25 \\
            $T_{\rm AB}$ (BJD$-243\,0000$) & 8703 & +57 & $-$68 \\
            $e_{\rm AB}$ & 0.044 & +0.041 & $-$0.030 \\
            $\omega_{\rm AB}$ ($^\circ$) & 22.7 & +6.5 & $-$7.1 \\
            $\gamma_{\rm AB}$ (km\,s$^{-1}$) & $-$13.95 & +0.84 & $-$0.88 \\
            $a_{\rm Aab}\sin i_{\rm A}$ (R$_\odot$)\tablefootmark{c} & 4.302 & +0.036 & $-$0.036 \\
            $q$ & 0.981 & +0.015 & $-$0.015 \\
            $M_{\rm Aa}\sin^3 i_{\rm A}$ (M$_\odot$) & 0.977 & +0.026 & $-$0.026 \\
            $M_{\rm Ab}\sin^3 i_{\rm A}$ (M$_\odot$) & 0.958 & +0.024 & $-$0.024 \\
            $a_{\rm A}\sin i_{\rm AB}$ (au)\tablefootmark{c} & 0.432 & +0.014 & $-$0.015 \\
            $f\!$($M_{\rm B}$) (M$_\odot$) & 0.000143 & +0.000014 & $-$0.000015 \\
            $M_{\rm B} \, (i_{\rm AB}\!=90^\circ$) (M$_\odot$) & 0.0850 & +0.0034 & $-$0.0033 \\
            $M_{\rm B} \, (i_{\rm AB}\!=90^\circ$) (M$_{\rm Jup}$) & 89.0 & +3.5 & $-$3.5 \\
            \hline 
        \end{tabular}
        }
        \tablefoot{ 
        \tablefoottext{a}{The values of $T_0$ and $P_{\rm A}$ were corrected by $c_0$ and $c_1$, respectively.\\}
        \tablefoottext{b}{The RV semi-amplitude $K_{\rm A}$ of the EB's centre of mass was computed from equation 19 of \citet{2024ApJ...976..242M}.\\}
        \tablefoottext{c}{For the eclipsing pair, we differentiate the semimajor axis $a_{\rm Aab}\!=a_{\rm Aa}\!+a_{\rm Ab}$ of the relative orbit from the semimajor axis $a_{\rm A}\!=a_{\rm AB}\!-a_{\rm B}$ of the barycentric orbit. Their respective inclinations are $i_{\rm A}\!=82.318^\circ$ (see Table~\ref{tab:MC_simu}) and $i_{\rm AB}\!=90^\circ$ (assumed).}
        }
    %\end{minipage}
\end{table*}

\subsection{Light-curve modelling}
\label{sec:LC_analysis}
%\textbf{Why PHOEBE}\\
Light-curve (LC) modelling for EBs allows determination of parameters such as stellar radii, temperature ratio, orbital period and inclination, that are extremely difficult to measure directly. There are numerous software codes available for modelling the light curves of EB systems \citep{1971ApJ...166..605W,2004MNRAS.349..547S, 2004MNRAS.351.1277S,2016A&A...591A.111M,2020ApJS..250...34C}, each with its unique advantages and limitations. We use version 4 of the \textsc{phoebe2} code\footnote{\url{http://phoebe-project.org/}} \citep{2016ApJS..227...29P,2018ApJS..237...26H, 2020ApJS..247...63J,2020ApJS..250...34C} to model the light curves. \textsc{phoebe2} adopts a complete treatment of Roche potential to accurately model the surface geometry of the stars combining it with other effects like limb darkening, gravity darkening and reflection. It also allows user to incorporate stellar spots which is a necessary feature to obtain a precise model for \RXGru{}.

%\textbf{Fitted parameters}
In order to reduce the degeneracy, it is beneficial to fix a few parameters to values obtained from a different analysis, while setting up the preliminary model of the system. We obtain the mass ratio ($q$), semimajor axis ($a$), and orbital period ($P$) from the combined RV+ETV solution (see Section~\ref{sec:rv_etv}), and fix $q$ and $a$ to these values. We manually set the logarithmic limb darkening coefficients to the values obtained from \citet{2017A&A...600A..30C}. In addition to the stellar and orbital parameters, we initialize the system with up to two spots (one on each component). 
%In addition to the stellar and orbital parameters, {\red\uwave{we initialize the system with one stellar spot on each component.}}
The initial parameterization defined by the longitude, co-latitude, radius, and relative temperature of the spots, is tweaked using forward models to visually match the out-of-eclipse shape of the light curve. Addition of more spots did not significantly improve the fit and hence we restrict ourselves to two spots in the \TESS{} LC solution and one spot in the \SWASP{} LC solution. In doing so, we assume that the hotter spot in \TESS{} is not visible in \SWASP{}.
%\tbf{{\red\uwave{We then added more spots to improve the fit. We stopped when additional spots did not significantly reduce the residuals. Hence we restricted ourselves to two spots in \TESS{}, one spot in \SWASP{}, and none in \Solaris{}. In doing so, we assumed that the hotter spot in \TESS{} is not visible in \SWASP{}.}}}

\begin{figure}
    \includegraphics[width=0.95\columnwidth]{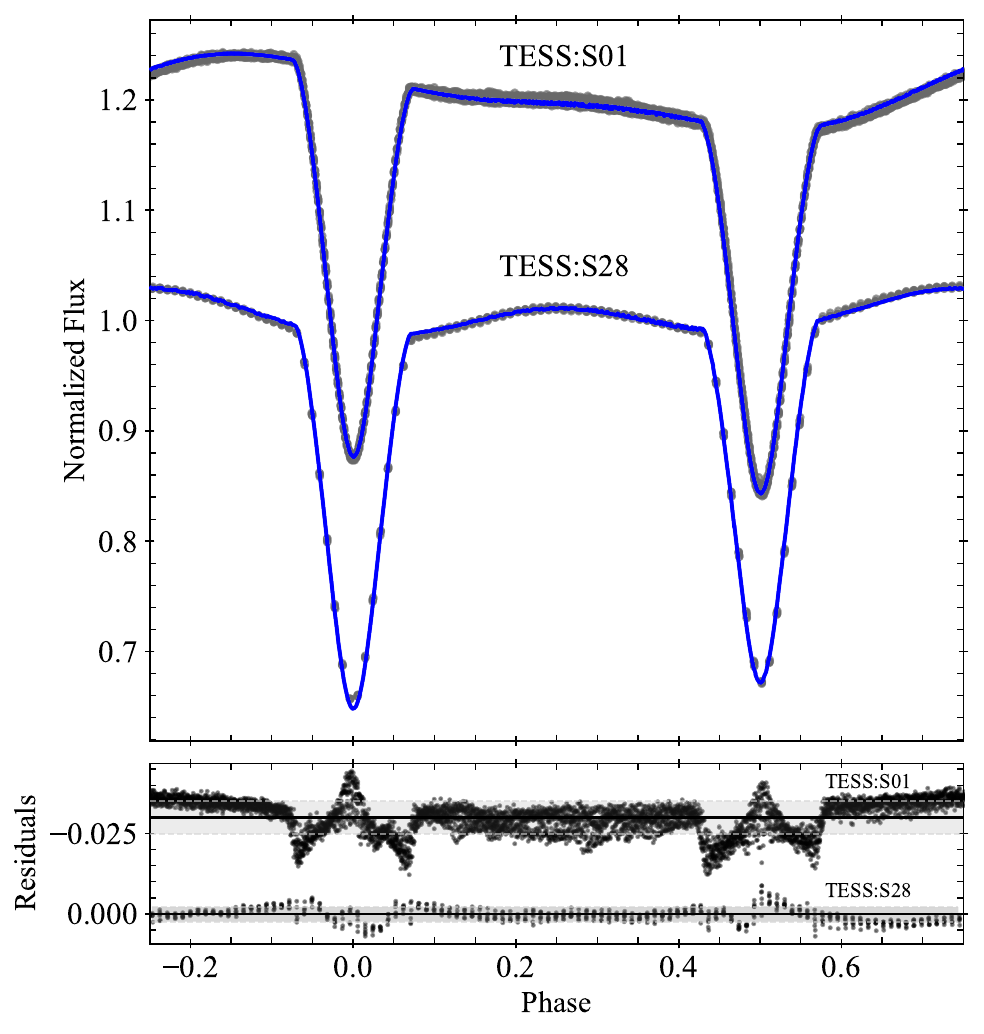}
    \caption{Best-fitting \textsc{phoebe2} model on phased \TESS{} LC observations for sector~1 and sector~28. The sector~1 light curve is shifted vertically for display purposes. The lower panel shows the corresponding residuals, with the shaded region indicating the 1$\sigma$ scatter.}
    \label{fig:s28tesslc}
\end{figure}

\begin{figure}
    \includegraphics[width=0.95\columnwidth]{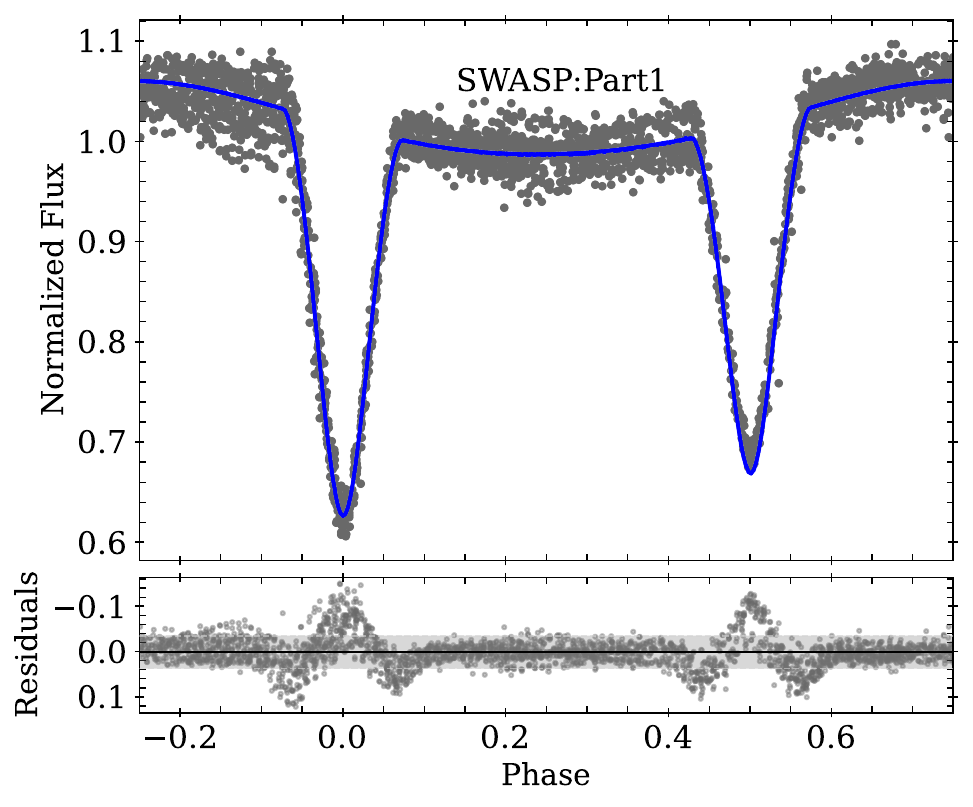}
    \caption{Best-fitting \textsc{phoebe2} model on phased \SWASP{} LC observations. The lower panel shows the corresponding residuals, with the shaded region indicating the 1$\sigma$ scatter.}
    \label{fig:swasplc}
\end{figure}

%\textbf{Optimization methods}\\
One of the key features of \textsc{phoebe2} (version 2.3 and higher) is a general framework to solve the inverse problem, i.e.\ obtaining stellar and orbital parameters based on the observational data. This is achieved using different optimization algorithms on a large set of forward models. For the case of \RXGru{}, we use the Nelder-Mead optimizer \citep{10.1093/comjnl/7.4.308} within \textsc{phoebe2.4}. The optimization is carried out in multiple steps. Initially, the spot parameters are kept fixed and other parameters, including stellar radii, time of superconjunction, passband luminosity, orbital period, temperature ratio and inclination are fitted. In the next step, we optimize for spot parameters like longitude and relative temperature, along with the temperature ratio and inclination of the system. All other stellar and orbital parameters are kept fixed to their previously optimized solution values. 
%The next step fixes some of these parameters to their optimized values and fits the spot parameters like longitude and relative temperature of the spot. 
A few passes of this step are implemented to ensure a robust set of parameters has been obtained. This process is performed independently for light curves from different sectors, since the eclipse depths and the overall shape of the out-of-eclipse regions change significantly due to the evolving spot. One can notice the effects of spot evolution on the out-of-eclipse shape of the light curves in Figs~\ref{fig:s28tesslc} and~\ref{fig:swasplc}. We still see substantial features in the \SWASP{} LC residuals near the eclipses. This can be explained by the fact that our averaged spot model does not account for brightness variations due to smaller spots, which become prominent when the spots are eclipsed.
%which are due to an averaged spot model as the variations due to smaller spots become prominent when they are eclipsed. 
The residuals can also be affected by our limb-darkening treatment.
%, but a more detailed analysis is beyond the scope of this paper.
%but an in-depth study is out of scope of this work. 
For this reason, we tested a range of limb-darkening coefficients with different temperature ratios,
%We rather try 
and tried to account for the deficiency in our spot modelling by sampling over some of the spot parameters when calculating uncertainties on binary parameters. Unfortunately, we were not able to improve the quality of the fit. This results in an underestimation of the fractional primary radius by about 7 per cent compared to the values derived from the analysis of the \TESS{} light curves. This discrepancy can be explained by the strong correlation between the radius ratio and the inclination. Thus, a higher value of $i_{\rm A}$ implies a higher value of $r_{\rm Ab}/r_{\rm Aa}$, and thus a smaller fractional primary radius, as evidenced in Table~\ref{tab:MC_simu}. Such a correlation was also noted by \citet{2020MNRAS.499.3019M} when comparing their results based on \TESS{} data with those obtained by \citet{2015MNRAS.448.1937C} for the multiple stellar system V1200~Cen using \SWASP{} data. Therefore, we decided to use only the results from the analysis of the high-precision \TESS{} photometric data to calculate the final values of the physical and orbital parameters  of the binary system.

%\textbf{Error estimation}\\
To ensure the uncertainties on the optimized parameters are robust and reliable, \textsc{phoebe2.4} implements MCMC sampling through \textsc{emcee} package \citep{2013PASP..125..306F, 2019JOSS....4.1864F}. This is a computationally demanding process, hence we switch to the \enquote{rotstar} model within \textsc{phoebe2.4} to construct models of the stars. This is much faster than the Roche model without compromising the accuracy of the parameters, when the stars are well separated and not too distorted.

We use 40 walkers to perform our sampling for a total of 12 parameters: period, inclination, eccentricity, argument of periastron, time of primary minima, radius ratio, sum of fractional radii, temperature ratio, longitude of the coldest spot, relative temperature of the spot, passband luminosity of the primary star, and third light. Ignoring the initial $\sim$1000 runs as the burn-in period, we let the walkers explore the parameter space for another 6000--7000 runs. This is done for each dataset (sector) individually. The final values of parameters and the errors obtained from the MCMC runs are mentioned in Table~\ref{tab:MC_simu}.\\

%\textbf{discuss the obtained parameters and mention the other parameters freed during MCMC}

%$L_s/L_p = 1.03_{-0.08 }^{+0.07}$

\begin{table*}
    \centering
    %\begin{minipage}{150mm}
        \caption{Parameters obtained from our analysis of the \TESS{}, \Solaris{}, and \SWASP{} light curves.}
        \label{tab:MC_simu}
        {
        \renewcommand{\arraystretch}{1.3}
        \begin{tabular}{@{}lccccc@{}} % four columns, alignment for each
        \hline\hline
        Parameter & \TESS{} S01 value & \TESS{} S28 value & \Solaris{} value & \SWASP{} value & Adopted value\tablefootmark{a} \\
        \hline
            $P_{\rm A}$ (d) & $0.743142_{-0.000015}^{+0.000014}$& $0.743162_{-0.000027}^{+0.000027}$& $0.743140122_{-0.000000061}^{+0.000000060}$ & $0.743119_{-0.000058}^{+0.000064}$& -- \\
            $T_0$ (BJD$-245\,0000$) &$8325.919732_{-0.000036}^{+0.000028}$ & $9075.7476_{-0.0027}^{+0.0027}$& $2061.99145_{-0.00045}^{+0.00047}$ &$3961.4537_{-0.0057}^{+0.0052}$ & -- \\
            %$T$ [BJD$-2\,450\,000$] & $8326.2496_{-0.000025}^{+0.000025}$ & $9075.5618_{-0.0027}^{+0.0027}$& $7061.8060_{-0.0005}^{+0.0005}$ & $3961.2680_{-0.0057}^{+0.0052}$& \\
            $e_{\rm A}$ & $0.00023_{-0.00018}^{+0.00029}$& $0.00107_{-0.00021}^{+0.00021}$& 0 (fixed) &$0.00034_{-0.00023}^{+0.00035}$ & -- \\
            $i_{\rm A}$ ($^\circ$) & $82.58_{-0.13}^{+0.13}$&$82.291_{-0.042}^{+0.042}$ & $83.49_{-0.13}^{+0.13}$ & $83.52_{-0.73}^{+0.86}$& $82.318 \pm 0.040$\\
            $r_{\rm Aa}$ & $0.2344_{-0.0057}^{+0.0058}$ & $0.2278_{-0.0068}^{+0.0086}$&$0.2168_{-0.0042}^{+0.0042}$ & $0.2165_{-0.0054}^{+0.0065}$& $0.2320 \pm 0.0046$\\
            $r_{\rm Ab}$ &$0.2349_{-0.0061}^{+0.0069}$ &$0.2373_{-0.0085}^{+0.0073}$ & $0.2318_{-0.0033 }^{+0.0033 }$& $0.2386_{-0.0061}^{+0.0051}$& $0.2359 \pm 0.0050$\\
            %$L_s/L_p$ &$x_{-0.0061}^{+0.0069}$ &$1.10_{-0.04}^{+0.04}$ & $1.03_{-0.08 }^{+0.07}$& $1.35_{-0.12}^{+0.11}$& \\
        \hline
        \end{tabular}
        }
        \tablefoot{
        %\tablefoottext{a}{The subscript A refers to the eclipsing pair, while the subscripts Aa and Ab refer to the primary and secondary EB components, respectively.\\}
        \tablefoottext{a}{The adopted values of $i_{\rm A}$, $r_{\rm Aa}$, and $r_{\rm Ab}$ were computed as the weighted mean of the \TESS{} values, which are less impacted by parameter correlations than the \Solaris{} and \SWASP{} values.}
        }
    %\end{minipage}
\end{table*}

\subsection{Spot modelling}
\label{sec:spots}

Since we have visible changes due to spots in the light curves, we tried to constrain the physical parameters of spots and their variation over time. \textsc{phoebe2.4} allows for adding and optimization of stellar spot parameters and has been used to check for their time evolution \citep{moharana2023}. Therefore, we derived the spot parameters: radius ($r^\mathrm{spot}$), relative temperature ($T^\mathrm{spot}$), co-latitude ($c^\mathrm{spot}$), and longitude ($l^\mathrm{spot}$) for RX~Gru from both the \SWASP{} and \TESS{} light curves. The parameters $c^\mathrm{spot}$ and $l^\mathrm{spot}$ are defined with respect to the star's reference frame, with $c^\mathrm{spot}$ equal to $0^\circ$ when the spot is at the north pole and $l^\mathrm{spot}$ equal to $0^\circ$ when the spot is facing the other star.
%\tbf{{\red\uwave{\boldm{$c^\mathrm{spot}$} and \boldm{$l^\mathrm{spot}$}} are defined relative to the \uwave{primary's configuration in the binary}. A \uwave{\boldm{$c^\mathrm{spot}$}} equal to zero means that the spot is at the north pole, and a \uwave{\boldm{$l^\mathrm{spot}$}} equal to zero means that the spot is facing the other star.}} 
For the spot modelling, we took two segments of \SWASP{} observations, each spread across 40$\,$d, and the first 15$\,$d of each \TESS{} sector. These specific segments were chosen as they have stable out-of-eclipse variations.
%These segments were relatively stable in terms of the out-of-eclipse variations.

\begin{figure}
    \includegraphics[width=\columnwidth]{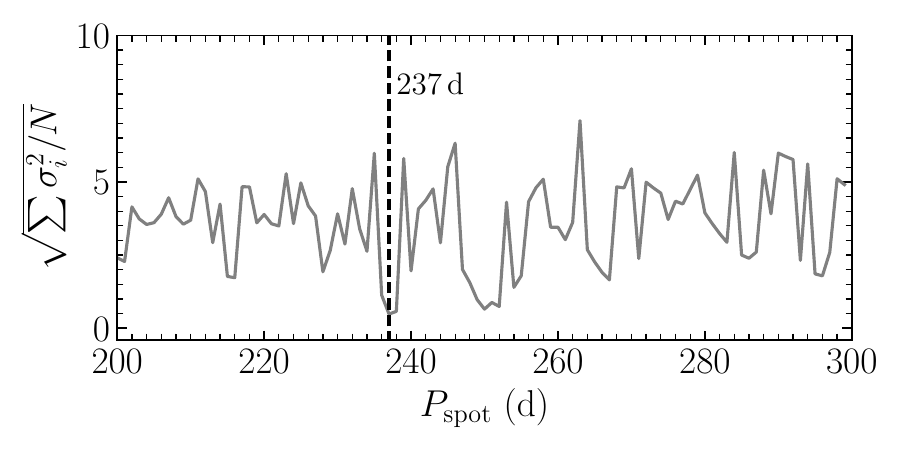}
    \includegraphics[width=\columnwidth]{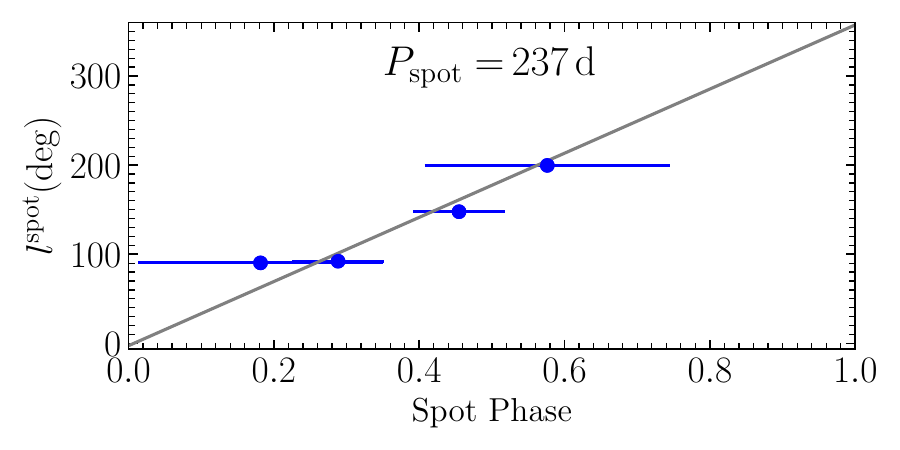}
    \caption{Top panel shows the rms residuals for different periods of $l^\mathrm{spot}$ variations. Bottom panel shows the phased variation of $l^\mathrm{spot}$ for the best-fitting spot migration period.}
    \label{fig:longvar}
\end{figure}

\begin{figure}
    \includegraphics[width=1.0\columnwidth]{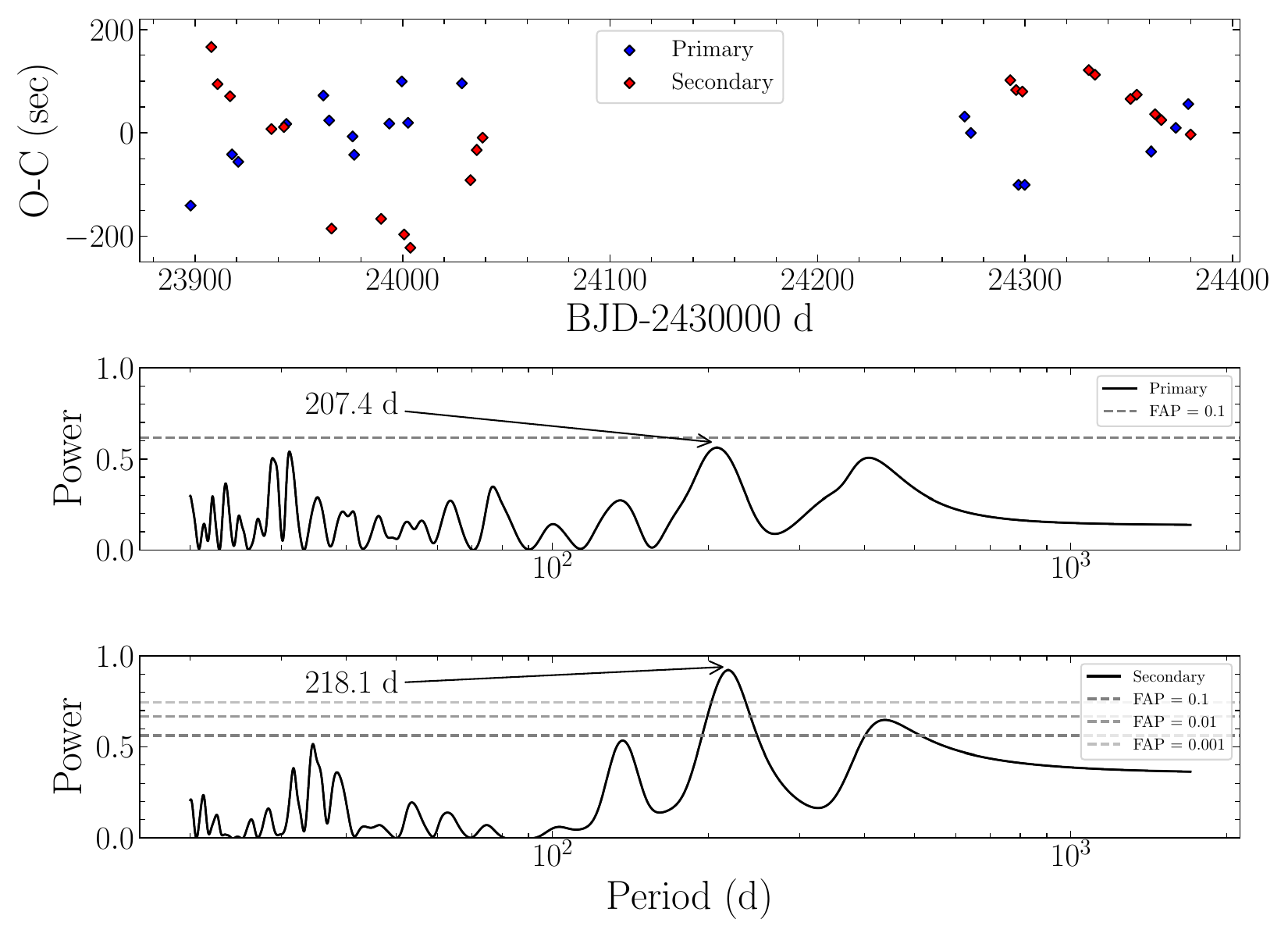}
    \caption{Top panel: $O-C$ diagram for the \SWASP{} primary and secondary eclipses after having removed the 23.8-yr modulation induced by the third body. Middle panel: LS periodogram of the primary ETV residuals. The horizontal dashed line depicts the FAP at 10 per cent. Bottom panel: same as the middle panel but for the secondary ETV residuals.}
    \label{fig:ETV_spots_periodogram}
\end{figure}

In Table~\ref{tab:coldspot}, we report for each segment the parameters of the coldest spot obtained from the spot modelling. This spot seems to be relatively stable, as can be seen from the similar values of $r^\mathrm{spot}$, $T^\mathrm{spot}$, and $c^\mathrm{spot}$. We also find that this spot migrates in longitude, as evidenced by the change in the $l^\mathrm{spot}$ value shown in Fig.~\ref{fig:longvar}. The longitude of a spot varies from 0 to $360^{\circ}$ during the spot migration period, $P_\mathrm{spot}$. Therefore, the $l^\mathrm{spot}$ variations will be along a straight line with slope equal to $360^{\circ}/P_\mathrm{spot}$. To search for the spot migration period, we calculated the rms of the residuals from a linear fit to $l^\mathrm{spot}$ assuming different periods from 100 to 2000$\,$d with a step size of 1$\,$d. The minimum rms value 
%$\sqrt{\sum{ \sigma_i^2 /N}}$
was found for a period of 237$\,$d. Additionally, it was shown by \cite{2013ApJ...774...81T} that migrating spots can produce ETVs with typical periods of the order of 50--200$\,$d and induce an anticorrelated behaviour of the primary and secondary minima variations. This behaviour is clearly seen in the ETV residuals of \RXGru{}, which are shown in the top panel of Fig.~\ref{fig:ETV_spots_periodogram}, especially during the interval of the \SWASP{} observations. We computed Lomb–Scargle (LS) periodograms \citep{1976Ap&SS..39..447L,1982ApJ...263..835S} of the primary and secondary ETV residuals and estimated the false alarm probability (FAP) of the maximum peak. Both periodograms exhibit a peak around 207--218$\,$d, with FAPs of $\sim$10 and $<$0.1 per cent, respectively. This value appears to be consistent with that obtained from the spot analysis. As can be seen in Fig.~\ref{fig:longvar}, there is a second minium at $\sim$250$\,$d that could correspond to the spot migration period. Unfortunately, the \SWASP{} observations cover only about one migration period, implying that we are not able to reliably discriminate between the two values. The values obtained from the periodogram analysis may also be affected by the limited coverage of the available observations. This is particularly true for the lowest period value of 207$\,$d due to the low amplitude of the primary ETV signal compared to the secondary one. This asymmetry can be explained by the relative visibility of the spot at a given orbital phase of the binary system (for more details, see \citealt{2013ApJ...774...81T}). For these reasons, we choose to provide only a range of possible values for $P_{\rm spot}$, comprised between $\sim$210 and 250$\,$d. A more detailed analysis, such as the one presented in \citet{2015MNRAS.448..429B}, would require longer light curves than currently available. Thus, the varying O’Connell effect seen in the light curve of \RXGru{} and the anticorrelated behaviour of the primary and secondary ETV curves can be explained by the presence of spots that migrate in longitude on the stellar surface due to differential rotation \citep{2015MNRAS.448..429B}. In the case of \RXGru{}, the short-term variability observed over the 15-yr period of observations suggests that the spots are long-lived, which is typical for young, rapidly rotating stars (\eg MML~48; \citealt{2025A&A...702A..17G}). We also highlight the case of KIC~12418816, a young, chromospherically active EB exhibiting similar ETV features as \RXGru{} (see fig.~9 of \citealt{2015MNRAS.448..429B}). For KIC~12418816, the spot migration period was estimated to be $\sim$260$\,$d by \citet{2018MNRAS.474..326D}, close to the range of values ($\sim$210--250$\,$d) obtained for \RXGru{}. This similarity tends to corroborate the activity-induced nature of the short-period signal observed in the $O-C$ diagram of \RXGru{}.

\begin{table*}
    \centering
    %\begin{minipage}{110mm}
        \caption{Spot parameters for the coldest spot on the secondary component of \RXGru{}. The parameters have been obtained for each segment using \textsc{phoebe2} modelling.}
        \label{tab:coldspot}
        {
        \renewcommand{\arraystretch}{1.0}
        \begin{tabular}{lcccc}
            \hline\hline
            Parameter & \SWASP{}~1 & \SWASP{}~2 & \TESS{} S01 & \TESS{} S28 \\
            \hline
            $T_\mathrm{seg}$ (BJD$-245\,0000$)\tablefootmark{a} & 4035.8572451  & 4366.3781826 & 8327.1474525 & 9077.5882681 \\
            $r^\mathrm{spot}$ (deg) & 37.6599& 37.6599 &38.0281 & 37.4141 \\
            $T^\mathrm{spot}$ &0.8145 & 0.9915 &0.8833 & 0.7574 \\
            $c^\mathrm{spot}$ (deg) & 30.0000 & 30.0000 &41.8299 & 11.1583 \\
            $l^\mathrm{spot}$ (deg) &90.2810 & 199.6614 &92.1624 & 147.6957  \\
            \hline
        \end{tabular}
        }
        \tablefoot{
        \tablefoottext{a}{$T_\mathrm{seg}$ corresponds to the mean time of each segment.}
        }
    %\end{minipage}
\end{table*}

%\begin{figure}
%    \includegraphics[trim = 2.5cm 1.93cm 3.0cm 2.0cm,clip,width=1.0\columnwidth,angle=0]{RXGru_residuals}
%    \caption{.}
%    \label{fig:ETV_spots}
%\end{figure}

\subsection{Spectral analysis}
\label{sec:specanalysis}

%\subsubsection{Spectral disentangling and \textsc{iSpec}}

To measure the atmospheric parameters of stars in the binary, we extracted individual spectra using spectral disentangling. We used the separation method implemented in the disentangling code \textsc{fd3} \citep{fdbinary} with a \textsc{python}-based wrapper\footnote{\url{https://github.com/ayushmoharana/fd3\_initiator}}. We selected a subset of the HARPS spectra where the line profiles were clearly separated for the two stars and used them for disentangling. The code takes stellar light fractions as inputs for each epoch, which were calculated using broadening functions (BFs; \citealt{bfsvd}). The code then optimizes the orbital elements of the binary orbit, initialized using the RV solutions, to separate the individual spectra. Using these spectra, the code calculates the corresponding residuals for every composite spectrum. We checked for any sharp features that could arise out of improper disentangling or static components like telluric and interstellar Na doublet lines. We did not find any such features and, therefore, we accepted the disentangling solution. The final spectra were cleaned for any RV offsets that could arise out of the disentangling routine.  We also corrected the continuum for any bias trends that propagate through the disentangling process \citep{fd3bias}. We used cubic splines to remove these trends. The final errors on the disentangled spectra are taken as the sum of the scaled residuals from the disentangling method and the S/N of the individual disentangled spectra. Details of the methodology can be found in \citet{2021MNRAS.508.5687H} and \citet{moharana2023}. 
%{\bf The accuracy of final solution could be checked by evaluating the epoch-wise residuals with respect to the final model. A proper disentangling would result in a uniform scatter and an improper disentangling would have features resembling spectral lines. We did not find any such feature, especially near  }

For our spectral analysis, we used \textsc{ispec} \citep{ispec2014,ispec2019}. Since disentangled spectra could be affected by biases, due to normalization or faint static lines, we avoided modelling individual lines and preferred an error-weighted joint fitting of all lines. We employed the synthetic spectral fitting (SSF) method, which consists of performing a $\chi^2$ minimization using grids of theoretical spectra that are synthesized on-the-fly. The SSF was carried out on the line lists from Gaia-ESO Survey (GES; \citealt{gso2012,gso2013}), which covers the wavelength range of 420--920\,nm. The SSF uses \textsc{spectrum}\footnote{\url{http://www.appstate.edu/~grayro/spectrum/spectrum.html}} to generate theoretical spectra with model atmospheres from \citet{marcges}. The solar abundances were chosen from \citet{grevesse2007}. 

\begin{figure}
    \includegraphics[width=\columnwidth]{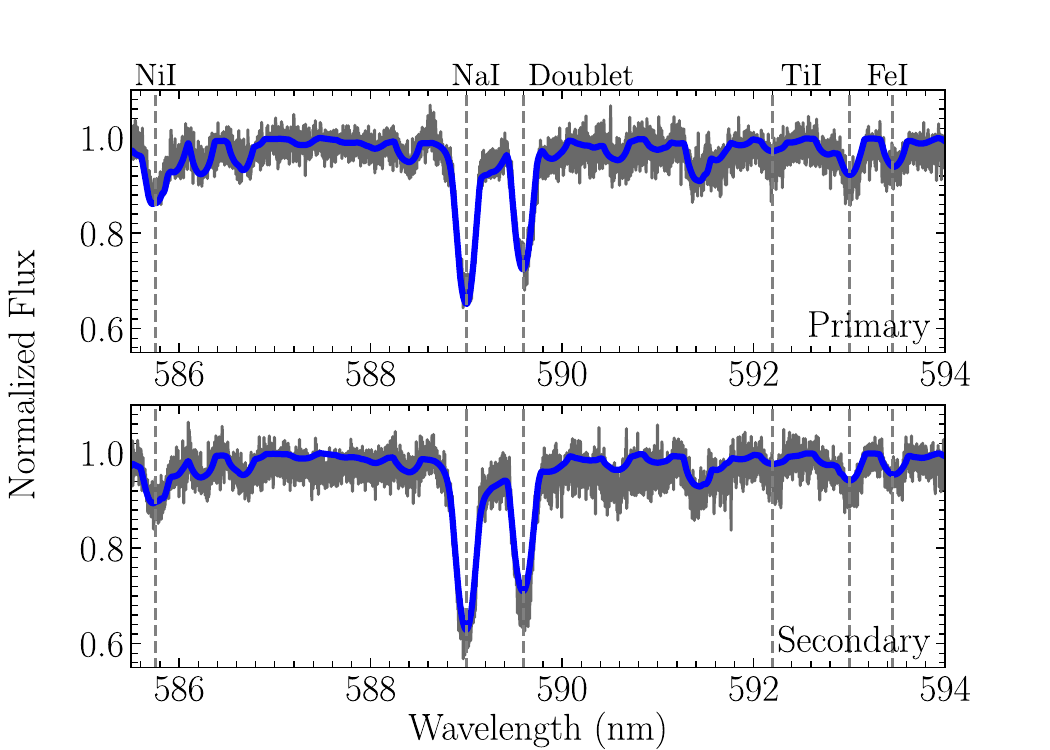}
    \includegraphics[width=\columnwidth]{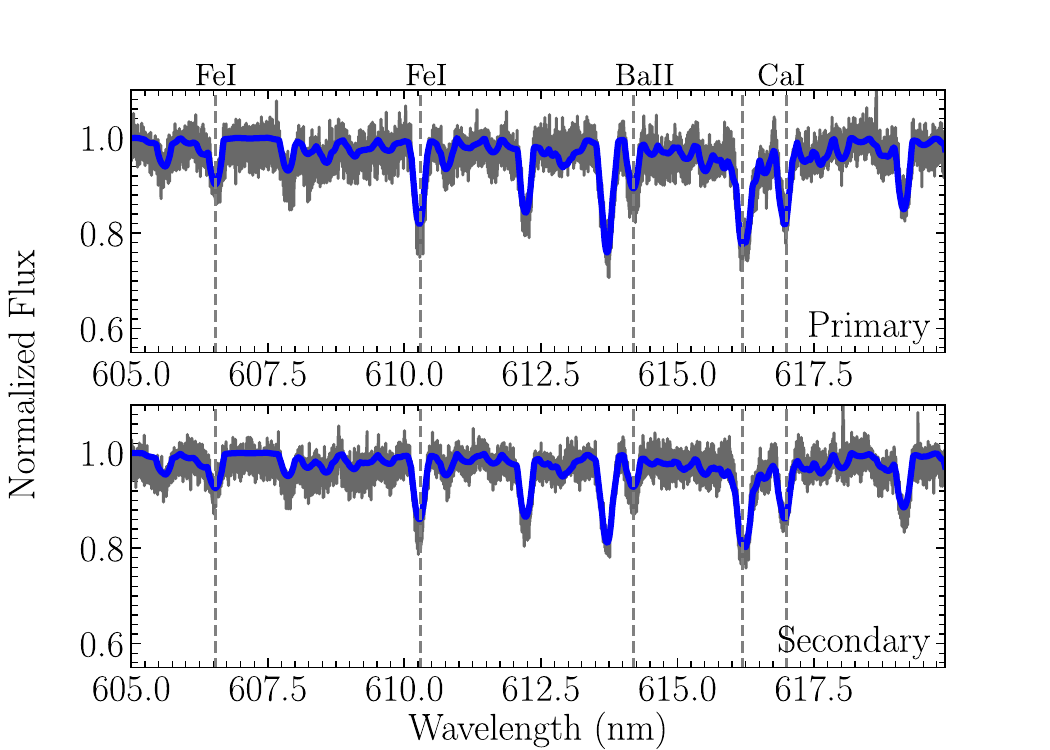}
    \caption{Model spectra (in blue) overplotted on the disentangled spectra (in grey) of the primary and secondary components.}
    \label{fig:spectra}
\end{figure}

We fit for effective temperature ($T_\mathrm{eff} $), metallicity ($\mathrm{[M/H]}_\mathrm{iSpec}$), helium over-abundance ($\alpha$), and micro-turbulence velocity ($v_\mathrm{mic}$). We fix the surface gravity ($\log g$) to the value obtained from the inferred mass and radius, while the projected rotational velocity ($v \sin i$) was taken from the BF analysis. The macro-turbulence velocity ($v_\mathrm{mac}$) was estimated in the fitting module using empirical relations. The limb-darkening coefficient (values were taken from \citealt{claretlimbdarkening}) and resolution were kept fixed  while $v_\mathrm{mac}$ was calculated automatically from an empirical relation established by GES and built in the code. The best-fitting synthetic spectra are shown in Fig.~\ref{fig:spectra}. The individual estimates for the primary and secondary stars are given in Table~\ref{tab:specanalysis}.

\begin{table}
    \centering
    %\begin{minipage}{71mm}
        \caption{Spectral analysis on disentangled spectra.}
        \label{tab:specanalysis}
        \begin{tabular}{lcc}
            \hline\hline 
            Parameter & Star Aa & Star Ab  \\
            \hline
            $T_\mathrm{eff} $ (K)  &$5379 \pm 289$ & $5322 \pm 278$  \\
            $\log g$ (dex) & 4.433\tablefootmark{a}&  4.411\tablefootmark{a}\\
            $\mathrm{[M/H]}_\mathrm{iSpec}$  (dex) &$-0.37 \pm 0.20$ & $-0.37 \pm 0.22$ \\
            $\alpha$  (dex)    &$0.27 \pm 0.18$&  $0.22 \pm 0.19$\\
            $v_\mathrm{mic}$ (km~s$^{-1}$) & $2.35 \pm 1.25$ & $1.40 \pm 1.27$ \\
            $v_\mathrm{mac}$ (km~s$^{-1}$)\tablefootmark{b}  & 3.54 & 3.48 \\
            $v \sin i$ (km~s$^{-1}$)& 66.99\tablefootmark{c} & 72.64\tablefootmark{c} \\
        \hline
    \end{tabular}
    \tablefoot{
    \tablefoottext{a}{From the inferred mass and radius.}
    \tablefoottext{b}{Obtained from empirical tables.}
    \tablefoottext{c}{Fixed from the BF analysis.}
    }
    %\end{minipage}
\end{table}

%%%%%%%%%%%%%%%%%%%%%%%%%%%%%%%%%%%%%%%%%%%%%%%%%%

%%%%%%%%%%%%%%%%% DISCUSSION %%%%%%%%%%%%%%%%%%%%%

\section{Discussion}
\label{sec:discussion}

\subsection{Physical parameters of \RXGru{}}

From our analysis of the light, RV, and ETV curves of \RXGru{}, we determined the stellar masses and radii of the EB components, as well as the minimum mass of the third body. In Table~\ref{tab:abs_param}, we presented the stellar parameters of each star Aa and Ab, along with their uncertainties, such as derived in this work.

We found that the eclipsing pair consists of two nearly equal-mass components in a circular, short-period ($P_{\rm A}=0.743\,$d) orbit. Although the measured values of $R$, $M$, and $T_{\rm eff}$ for both stars overlap within their errors, our analysis suggests that the hotter primary component is the smaller and more massive one. The derivation of the effective temperature will be discussed in detail in Section~\ref{sec:teff}. For the stellar mass and radius, we reached a precision better than 2.7 per cent. In the calculations, the inner eccentricity was fixed to the value determined from the fit to the ETV and RV curves ($e_{\rm A} \simeq 0.003$), rather than to the values derived from the LC modelling of the \TESS{} and \SWASP{} data, as the latter may be affected by activity-induced variability. This does not change the values of the component masses within the 1$\sigma$ uncertainties. For the stellar radii, we adopted the weighted mean of the \TESS{} values, as reported in Table~\ref{tab:MC_simu}. The reader will note the small scatter in the \TESS{} LC residuals (see Fig.~\ref{fig:s28tesslc}). We also point out the good agreement between the values obtained from the \TESS{} sectors~1 and~28 light curves. By modelling the two data sets independently, we mitigate the impact of the evolution of spots on the derived stellar radii, as demonstrated by \citet{2021MNRAS.508.5687H}. For star Aa, we obtained $M_{\rm Aa} = 1.004^{+0.027}_{-0.026}\,$M$_\odot$ and $R_{\rm Aa} = 1.007\pm0.021\,$R$_\odot$, while for star Ab, we obtained $M_{\rm Ab} = 0.985^{+0.024}_{-0.025}\,$M$_\odot$ and $R_{\rm Ab} = 1.024\pm0.023\,$R$_\odot$. These results will be compared with predictions of stellar models in Section~\ref{sec:models}.

In principle, the mass of the third body can be obtained from Kepler's third law applied to the EB's barycentric orbit, namely
\begin{equation}
\frac{M_{\rm B}^3}{(M_{\rm A}+M_{\rm B})^2} = \frac{a_{\rm A}^3}{P_{\rm AB}^2},
\end{equation}
where $M_{\rm A} = M_{\rm Aa} + M_{\rm Ab}$ is the sum of the masses of the eclipsing components Aa and Ab. Here, the projected semimajor axis $a_{\rm A}\sin i_{\rm AB}$ is related to $A_{\rm LTTE}$ by equation~(\ref{eq:Altte}). We can then write the mass function of the third body as
\begin{equation}\label{eq:fm}
    \begin{split}
        f(M_{\rm B}) & = \frac{M_{\rm B}^3 \, \sin^3 i_{\rm AB}}{(M_{\rm Aa}+M_{\rm Ab}+M_{\rm B})^2} \\
        & = 1.074 \times 10^{-3} \, \frac{A_{\rm LTTE}^3}{P_{\rm AB}^2},
    \end{split}
\end{equation}
where the masses are expressed in the units of solar mass, $A_{\rm LTTE}$ is in seconds, and $P_{\rm AB}$ is in days. Assuming an inclination of $i_{\rm AB}\!=90^\circ$, equation~(\ref{eq:fm}) provides a lower limit on the mass $M_{\rm B}$ of the third component. We solved the mass function for the minimum tertiary mass by means of Laguerre's method \citep{1992nrca.book.....P}, using the posterior distribution samples of $M_{\rm Aa}$, $M_{\rm Ab}$, $A_{\rm LTTE}$, and $P_{\rm AB}$ obtained from our MCMC fitting in Section~\ref{sec:rv_etv}. We then computed the median and the 16 per cent and 84 per cent credible intervals on the minimum tertiary mass from the corresponding posterior probability distribution. We obtained $M_{\rm B} = 89.0\pm3.5\,$M$_{\rm Jup}$ (see Table~\ref{tab:RV_param}), which is just above the hydrogen-burning mass limit of $\sim$80$\,{\rm M}_{\rm Jup}$ that separates brown dwarfs from very low-mass stars \citep{2002A&A...382..563B}. The tertiary companion is thus a massive brown dwarf or a very low-mass star orbiting the close eclipsing pair with a relatively long period of $\sim$23.8$\,$yr. The implications of these results will be discussed in detail in Section~\ref{sec:evolution}.

\begin{table}
    \centering
    %\begin{minipage}{71mm}
        \caption{Stellar parameters and distance of \RXGru{}.}
        \label{tab:abs_param}
        {
        \renewcommand{\arraystretch}{1.0}
        \begin{tabular}{@{}lccc@{}}
            \hline\hline
            & & 84 per cent & 16 per cent \\
            Parameter & Median & interval & interval \\
            \hline
            $a_{\rm Aab}$ (R$_\odot$) & 4.341 & +0.036 & $-$0.036 \\
            $M_{\rm Aa}$ (M$_\odot$) & 1.004 & +0.027 & $-$0.026 \\
            $M_{\rm Ab}$ (M$_\odot$) & 0.985 & +0.024 & $-$0.025 \\
            $R_{\rm Aa}$ (R$_\odot$) & 1.007 & +0.021 & $-$0.021 \\
            $R_{\rm Ab}$ (R$_\odot$) & 1.024 & +0.023 & $-$0.023 \\
            $\log g_{\rm Aa}$ & 4.433 & +0.018 & $-$0.017 \\
            $\log g_{\rm Ab}$ & 4.411 & +0.019 & $-$0.018 \\
            $T_{\rm eff,\,Aa}$ (K) & 5379\tablefootmark{a} & \multicolumn{2}{c}{$\pm$289} \\
            $T_{\rm eff,\,Ab}$ (K) & 5322\tablefootmark{a} & \multicolumn{2}{c}{$\pm$278} \\
            $T_{\rm eff,\,Ab}/T_{\rm eff,\,Aa}$ & 0.989\tablefootmark{a} & \multicolumn{2}{c}{$\pm$0.075} \\
            $L_{\rm Aa}$ [$\log\,(L$/L$_\odot)$] & $-$0.118 & +0.090 & $-$0.094 \\
            $L_{\rm Ab}$ [$\log\,(L$/L$_\odot)$] & $-$0.122 & +0.088 & $-$0.092 \\
            $d$ (pc) & $163.2$ & $+12.6$ & $-12.0$ \\
            $\pi$ (mas) & $6.13$ & $+0.49$ & $-0.44$ \\
            \hline
        \end{tabular}
        }
        \tablefoot{
        \tablefoottext{a}{From the spectral analysis presented in Section~\ref{sec:specanalysis} (see also Table~\ref{tab:specanalysis}).}
        }
    %\end{minipage}
\end{table}

\subsection{Effective temperature and distance}
\label{sec:teff}

In Section~\ref{sec:specanalysis}, we derived the effective temperatures of the two eclipsing components from the analysis of HARPS spectra, which, when combined with the apparent visual magnitude and stellar radii, allow us to determine the distance of the system.

In our calculations, we used the bolometric correction (BC) tables\footnote{\url{https://github.com/casaluca/bolometric-corrections}} from \citet{2018MNRAS.479L.102C,2018MNRAS.475.5023C} and adopted solar values $T_{{\rm eff,}\odot}=5777\,$K and $M_{{\rm bol,}\odot}=4.74\,$mag. By considering the values of $T_{\rm eff}$, $\log g$ and [M$/$H], listed in Table~\ref{tab:specanalysis}, and an interstellar reddening $E(B - V) =$ $0.05\,$mag \citep{2018yCat.2354....0G}, we obtained ${\rm BC}_{\rm Aa} =$ $-0.310\,$mag and ${\rm BC}_{\rm Ab} =$ $-0.324\,$mag. From the apparent visual magnitude of the system, $V_{\rm syst} = 10.658\pm0.030\,$mag, we then derived a photometric parallax of $6.13^{+0.49}_{-0.44}\,$mas for \RXGru{}. This value can be directly compared with the trigonometric parallax from the \Gaia{} Data Release~3 (DR3; \citealt{2022yCat.1355....0G}), namely $\pi = 6.935\pm0.017\,$mas. We note that our parallax estimate does not match the new \Gaia{} DR3 value within their respective error bars. Such a discrepancy was also found for some other triple systems (\eg TIC~337993842; \citealt{2025A&A...703A.153B}), despite having low RUWE values. In order to check the derived $T_{\rm eff}$ values, we constructed an SED with photometry queried from the SIMBAD database \citep{2000A&AS..143....9W}. For this, we used the \textsc{SEDfit}\footnote{\url{https://github.com/mkounkel/SEDFit}} code \citep{sedfit}, which allows fitting a multiple-star SED assuming a single metallicity for the whole system. Thus, we fitted a binary SED with the stellar parameters fixed, only varying the distance $d$, the interstellar extinction $A_{\rm V}$, and the metallicity [Fe/H]. A comparison between the observed SED and the best-fitting model is shown in Fig.~\ref{fig:SED}, while the corresponding flux measurements are provided in Table~\ref{tab:sed_phot} in Appendix~\ref{app:sed}. As discussed later in Section~\ref{sec:evolution}, the observed SED reveals a UV excess consistent with the picture of two young stars undergoing shared accretion \citep{2017A&A...599A..27G}. From the best-fitting model, we obtained the following results: $d = 165.8\,$pc, $A_{\rm V}= 0.25\,$mag, and ${\rm[Fe/H]} = -0.49$. These values agree well with those reported above, thereby reinforcing the reliability of our $T_{\rm eff}$ estimates.

For comparison, we searched for the effective temperatures that best match the \Gaia{} DR3 parallax. We assumed the same effective temperature ratio as derived from the spectral analysis and obtained $T_{\rm eff,Aa} =$ $5147\pm277\,$K and $T_{\rm eff,Ab} =$ $5093\pm266\,$K (see Table~\ref{tab:teff}). The corresponding BCs are found to be $-0.376$ and $-0.395\,$mag for stars Aa and Ab, respectively. We point out that in the case of \RXGru{}, the parallax measurement from \Gaia{} DR3 could be affected by the orbital motion of the system (see \eg \citealt{2008IAUS..248...59P}) and thus lead to incorrect estimation of the effective temperatures. Indeed, we did not find solutions for \RXGru{} in the \Gaia{} DR3 non-single-stars (NSS) catalogue \citep{2022yCat.1357....0G}, implying that the system was treated as a single star. Therefore, we decided to use the $T_{\rm eff}$ values derived from the spectral analysis for the subsequent modelling of the two eclipsing components of \RXGru{}.

\begin{figure}
    \centering
    \includegraphics[width=\columnwidth]{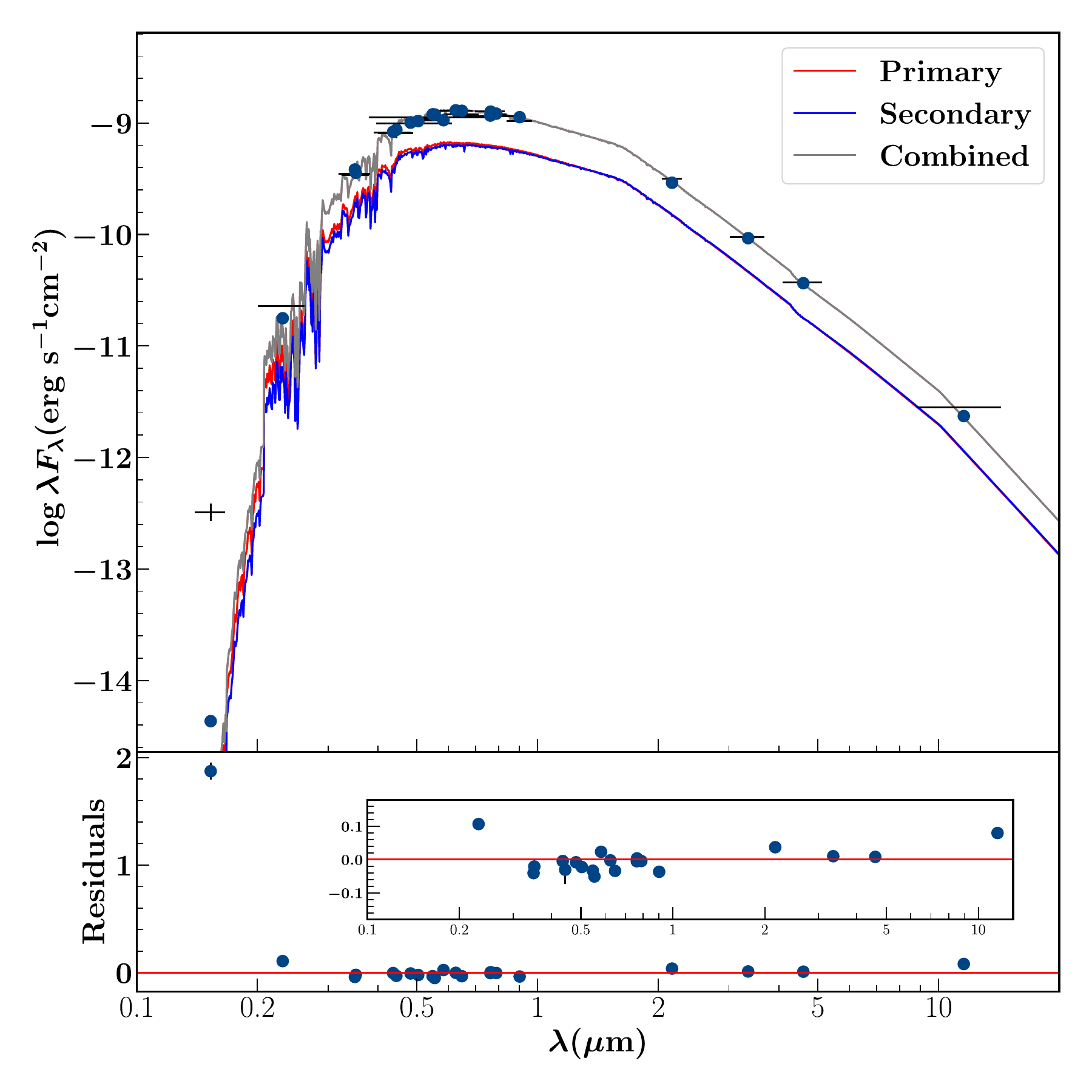}
    \caption{Best-fitting SED for \RXGru. The upper panel shows the combined binary fluxes in dark blue circles against the observed fluxes in crosses with the length of the crosses representing errors on flux and the width of the filter. The combined binary SED is represented in grey while the primary and secondary star SEDs are in red and blue, respectively. The lower panel shows the residuals from the fit. We see a UV excess in the GALEX fluxes with the GALEX FUV showing a strong discrepancy from the model.}
    \label{fig:SED}
\end{figure}

%\begin{table}
%    \centering
%        \caption{\tbf{Caption.}}
%        \label{tab:SEDfit}
%        \begin{tabular}{ccc}
%            \hline
%            Parameter & Primary & Secondary  \\
%            \hline
%            \multicolumn{3}{l}{Binary Fixed} \\
%            \hline
%            Radius (R$_{\odot}$) & 1.007  & 1.024 \\
%            T$_\mathrm{eff}$ (K) & 5422  & 5322 \\
%            $\log{g}$ (dex) & 4.43  & 4.41 \\
%            Mass (M$_{\odot}$) & 0.96  & 0.95  \\
%            Distance (pc) & \multicolumn{2}{c}{165.755} \\
%            $A_v$ (mag) & \multicolumn{2}{c}{0.246} \\
%            $[$Fe/H$]$ (dex) & \multicolumn{2}{c}{$-$0.49} \\
%            $\chi^2$ & \multicolumn{2}{c}{46.7} \\
%            \hline
%        \end{tabular}
%\end{table}

\begin{table}
    \centering
    %\begin{minipage}{47mm}
        \caption{Effective temperatures and luminosities of \RXGru{} computed using the \Gaia{} DR3 parallax.}
        \label{tab:teff}
        {
        \renewcommand{\arraystretch}{1.0}
        \begin{tabular}{@{}lcc@{}}
            \hline\hline
            Parameter & Value & 1$\sigma$ error \\
            \hline
            $T_{\rm eff,\,Aa}$ (K)\tablefootmark{a} & $5147$ & $277$ \\
            $T_{\rm eff,\,Ab}$ (K)\tablefootmark{a} & $5093$ & $266$ \\
            $L_{\rm Aa}$ [$\log\,(L$/L$_\odot)$] & $-0.197$ & 0.096 \\
            $L_{\rm Ab}$ [$\log\,(L$/L$_\odot)$] & $-0.201$ & 0.093 \\
            \hline
        \end{tabular}
        }
        \tablefoot{
        \tablefoottext{a}{Assuming the same effective temperature ratio and relative uncertainties as derived from the spectral analysis.}
        }
    %\end{minipage}
\end{table}

\subsection{Comparison with stellar models}
\label{sec:models}

This section is dedicated to the comparison between the results from our spectral and orbital analysis of \RXGru{} and the theoretical predictions from stellar models. The age determination of each of the two eclipsing stars will then help us to shed light on the evolutionary status of \RXGru{}.

\subsubsection{\MESA{} isochrones}

In order to determine the age of the two stars Aa and Ab, we generated a set of isochrones using a dedicated web interface\footnote{\url{http://waps.cfa.harvard.edu/MIST/}} based on the Modules for Experiments in Stellar Astrophysics (\MESA{}; \citealt{2011ApJS..192....3P,2013ApJS..208....4P,2015ApJS..220...15P,2018ApJS..234...34P}) and developed as part of the \MESA{} Isochrones and Stellar Tracks project (MIST v1.2; \citealt{2016ApJ...823..102C,2016ApJS..222....8D}). We considered in this work only the case of non-rotating stars ($v/v_{\rm crit} = 0$). For both stars, we adopted the solar mixture from \citet{2009ARA&A..47..481A}, which corresponds to $Y_{\odot{\rm ,ini}}=0.2703$ and $Z_{\odot{\rm ,ini}}=0.0142$. We then searched for the isochrone that best matches the observed parameters ($R$, $M$, and $T_{\rm eff}$) of each star. We adopted the effective temperatures derived in Section~\ref{sec:specanalysis} and provided in Table~\ref{tab:specanalysis}.

\begin{figure}
    \includegraphics[trim = 0.5cm 0.0cm 1.0cm 1.0cm,clip,width=1.0\columnwidth,angle=0]{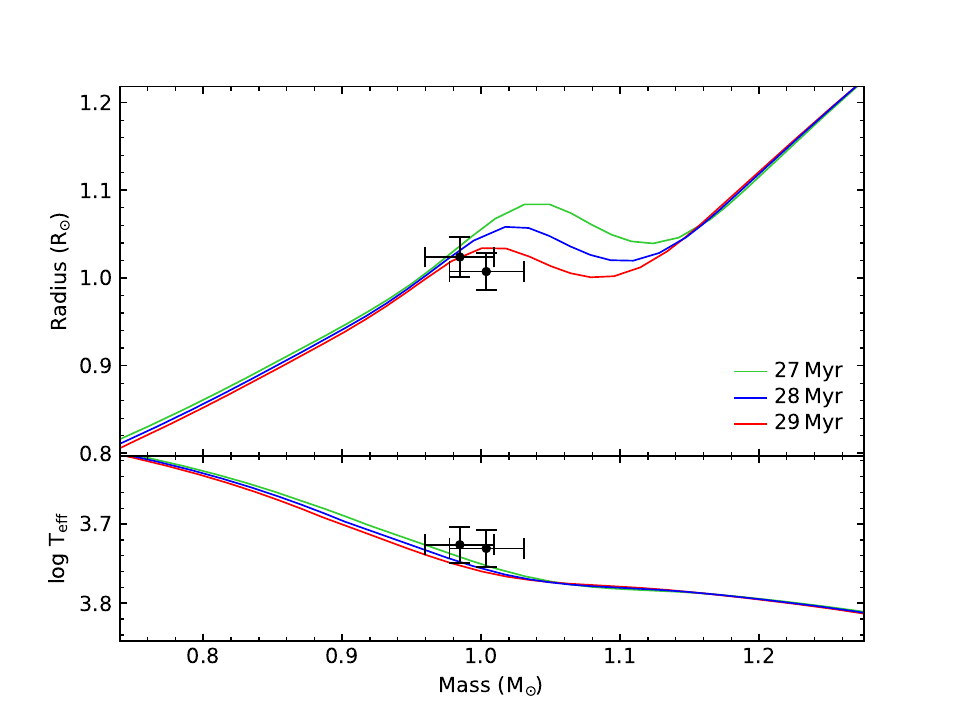}
    \caption{Comparison between the observed parameters of \RXGru{} and the predictions from \MESA{} isochrones. Green, blue, and red lines correspond to isochrones for ages of 27, 28, and 29$\,$Myr, respectively. Black dots with error bars indicate the derived values of $R$, $M$, and $\log\,T_{\rm eff}$ with their corresponding 1$\sigma$ uncertainties. The $T_{\rm eff}$ values are taken from Table~\ref{tab:specanalysis}.}
    \label{fig:mesa_iso}
\end{figure}

The comparison between the observed parameters from our analysis of \RXGru{} and the predictions from \MESA{} isochrones is shown in Fig.~\ref{fig:mesa_iso}. For both stars, we found that the parameters $R$, $M$, and $T_{\rm eff}$ match well the 27--29$\,$Myr isochrones within their 1$\sigma$ error bars, assuming a solar metallicity. We also investigated the effect of metallicity on the age determination. In particular, the $T_{\rm eff}$ values derived from the \Gaia{} DR3 parallax can be reproduced well by a 29-Myr isochrone for a higher metallicity of [Fe$/$H]$_{\rm ini}=0.22$ (\ie $Y_{\rm ini}=0.2833$ and $Z_{\rm ini}=0.0229$), which is not consistent with the metallicity derived from our spectral analysis. Overall, the best agreement was found by adopting the solar metallicity and the effective temperatures of $T_{\rm eff,Aa} = 5379\pm289\,$K and $T_{\rm eff,Ab} = 5322\pm278\,$K (see Table~\ref{tab:specanalysis}). We conclude that with a common age of $\sim$28$\,$Myr, both eclipsing components of \RXGru{} have almost reached the end of the PMS. These results will be compared with those from another evolutionary code described in the next section.

\subsubsection{Cesam2k20 stellar models}

Models were computed using the Cesam2k20 stellar evolution code\footnote{Cesam2k20 is freely available at \url{https://www.ias.u-psud.fr/cesam2k20/home.html}.} \citep[see][] {1997A&AS..124..597M,2008Ap&SS.316...61M,2013A&A...549A..74M,2025arXiv251102801M}. We have used the OPAL equation of state \citep{2002ApJ...576.1064R} and opacities \citep{1996ApJ...464..943I}, complemented at $T < 10^4\,$K by the WSU low-temperature opacities \citep{2005ApJ...623..585F}. We used the NACRE II nuclear reaction rates \citep{2013NuPhA.918...61X}, and the initial solar composition of \cite{2009ARA&A..47..481A}. The atmosphere is computed in the gray approximation of Eddington and integrated up to an optical depth of $\tau = 10^{-4}$. The temperature gradient in convective zones was computed using the mixing-length theory \citep[e.g.][]{1958ZA.....46..108B} with a mixing length $l=\alpha H_{\rm P}$. The initial chemical composition is that of a calibrated solar model using diffusion: $Y_{\rm ini} = 0.2621$ and $Z_{\rm ini} = 0.01477$. We used the mixing-length parameter $\alpha$ that calibrates the same solar model: $\alpha = 1.7605$.

\begin{figure}
    \includegraphics[trim = 0.1cm 0.1cm 1.5cm 1.3cm,clip,width=1.0\columnwidth,angle=0]{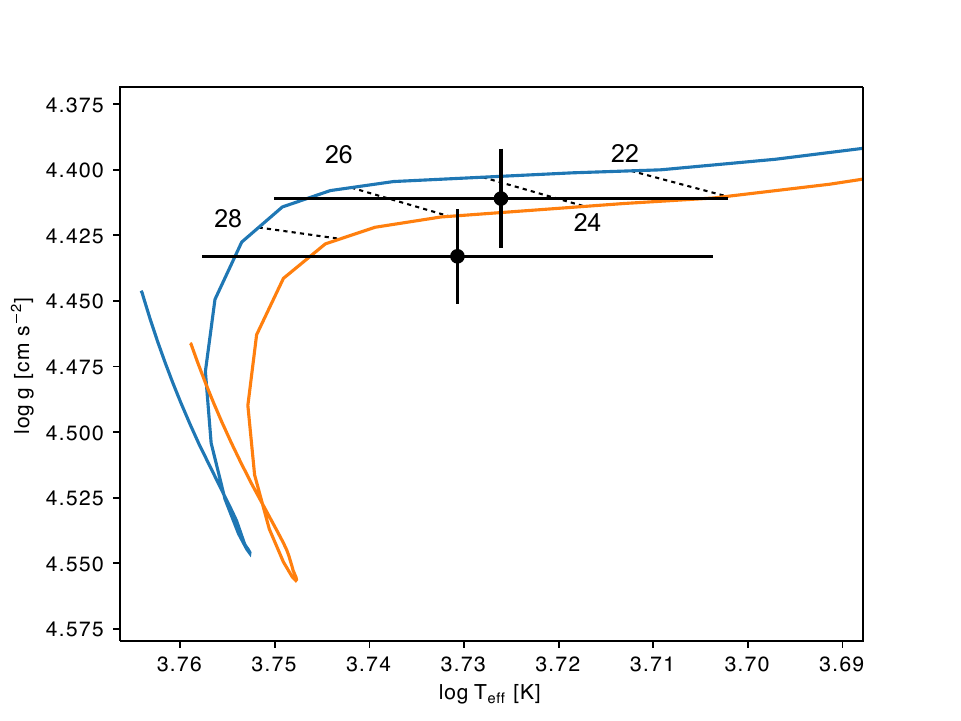}
    \caption{Evolutionary tracks calculated with Cesam2k20. Both stars are at the final stages of the PMS.}
    \label{fig:CResults}
\end{figure}

Fig.~\ref{fig:CResults} shows the evolutionary tracks of both components (with masses 1.004 and $0.985\,$M$_{\odot}$) on the Kiel diagram. Both components are fitted at an age between 23 and 28 Myr, showing that both stars are at the final stages of the PMS phase. We computed models with lower and higher metallicities, but they fail to reproduce the observed $\log g$ and $T_{\rm eff}$ at any age. These results are consistent with the previous section.

%\begin{figure}
%    \includegraphics[width=1.0\columnwidth,angle=0]{Figure_2}
%    \caption{Evolutionary tracks calculated with Cesam2k20}
%    \label{fig:CResults}
%\end{figure}

\subsubsection{Evolutionary status of \RXGru{}}
\label{sec:evolution}

Using two different stellar models, we determined the age of the \RXGru{} system from the observed parameters of its eclipsing components. We obtained a good agreement between \MESA{} and Cesam2k20 models, which predict an age of $\sim$28{\,}Myr for both components. From our orbital analysis, we also identified a tertiary companion (a massive brown dwarf or a very low-mass star) in orbit around the tight inner binary with a relatively long period of $\sim$23.8$\,$yr. So far, only a few PMS triples with well-defined orbital periods have been described in the literature. These are RS~Cha \citep{2013MNRAS.432..327W}, TY~CrA \citep{1996A&A...310..228C}, MML~53 \citep{2019A&A...623A..23G}, V1200~Cen \citep{2015MNRAS.448.1937C}\footnote{V1200~Cen was later identified as a quadruple system with a 180-d outer period by \citet{2020MNRAS.499.3019M}.}, GW~Ori \citep{2017ApJ...851..132C}, TWA~3 \citep{2017ApJ...844..168K}, V807~Tau \citep{2012ApJ...756..120S}, and TIC~167692429 and TIC~220397947 \citep{2020MNRAS.493.5005B}. As noticed by \citet{2020ApJ...902..107L}, all of these systems are found to have an outer-to-inner period ratio lower than $10^{3.5}$ (see their fig.~7), implying a relatively tight configuration. It is worth noting that \RXGru{} has the highest outer-to-inner period ratio ($\sim$10$^{4.1}$) among the aforementioned systems, making it an extremely interesting object for studying the early formation processes of close binaries in triple systems with relatively wide outer orbits.

Based on our findings, we can derive a global picture of the evolutionary status of \RXGru{}. This system was likely formed $\sim$28{\,}Myr ago in an unstable non-hierarchical configuration. After a series of chaotic interactions, the system dynamically unfolds into a more stable hierarchical structure on a time-scale of 1--10$\,$Myr, leaving the tertiary component in a wide orbit around the inner binary \citep{2012Natur.492..221R}. As explained by these authors, the presence of a gravitational potential, such as that of the nascent cloud core, is needed in order to ensure the long-term stability of the system. This mechanism allows the formation of triple stellar systems with a wide range of tertiary-to-binary mass ratios, depending on their dynamical and accretion history (see fig.~3 of \citealt{2012Natur.492..221R}). At the same time, the inner binary is expected to interact with the circumbinary material during the accretion phase, leading to the shrinkage of the binary orbit, due to energy dissipation, and to the equalization of the component masses \citep{2000MNRAS.314...33B,2000A&A...360..997T}. In their numerical simulations of triple-star dynamical evolution, \citet{2018ApJ...854...44M} found that $\sim$60 per cent of close binaries with orbital periods $<\,$10$\,$d form in this manner. In addition, by re-analysing the binary statistics of the \citet{2010ApJS..190....1R} sample, which was derived from a survey of stellar multiplicity in the solar neighbourhood, \citet{2017ApJS..230...15M} estimated that $\sim$20--30 per cent of solar-type binaries with orbital periods $<\,$20$\,$d have a secondary-to-primary mass ratio greater than 0.9. These results support our conclusions regarding the evolutionary status of \RXGru{}.

%%%%%%%%%%%%%%%%%%%%%%%%%%%%%%%%%%%%%%%%%%%%%%%%%%

%%%%%%%%%%%%%%%%% SUMMARY %%%%%%%%%%%%%%%%%%%%%%%%
 
\section{Summary}
\label{sec:summary}

In this work, we reported the discovery of a new short-period PMS EB, \RXGru{}, orbited by a distant circumbinary companion. For this, we made use of the \Solaris{}, \TESS{}, and \SWASP{} photometry, as well as the new and archival RV measurements from four different high-resolution spectrographs, namely HARPS, FEROS, CHIRON, and HRS. Despite the presence of migrating spots on the stellar surface, our analysis of the light curves, combined with the RV measurements, allowed us to derive the mass and radius of each eclipsing component with a precision better than 2.7 per cent. In particular, we found the eclipsing pair to be composed of two twin components with nearly equal masses and radii, \ie $M_{\rm Aa} = 1.004^{+0.027}_{-0.026}\,$M$_\odot$ and $R_{\rm Aa} = 1.007\pm0.021\,$R$_\odot$ for star Aa, and $M_{\rm Ab} = 0.985^{+0.024}_{-0.025}\,$M$_\odot$ and $R_{\rm Ab} = 1.024\pm0.023\,$R$_\odot$ for star Ab. Based on our analysis of HARPS spectra, we also derived the $T_{\rm eff}$ values of the two eclipsing components, which are found to be $T_{\rm eff,Aa} = 5379\pm289\,$K and $T_{\rm eff,Ab} = 5322\pm278\,$K. We used these values, combined with the apparent visual magnitude and stellar radii, to estimate the photometric parallax of the system, $\pi =$ $6.13^{+0.49}_{-0.44}\,$mas. Our parallax estimate appears to be in strong disagreement with the new \Gaia{} DR3 value. However, we point out that \RXGru{} was treated as a single star during the \Gaia{} DR3 processing, resulting in potential biases in the parallax determination. Additionally, we compared the observed parameters of each eclipsing component with the predictions from two independent stellar evolution codes, namely \MESA{} and Cesam2k20. Both codes give consistent results, and are able to reproduce the observed parameters ($R$, $M$, and $T_{\rm eff}$) for a solar metallicity and a common age of $\sim$28{\,}Myr. This places the two stars at the very end of the PMS phase.

Finally, we detected a circumbinary companion around the EB by applying the ETV method to the \Solaris{}, \TESS{}, and \SWASP{} data. We found the minimum tertiary mass to be $89.0\pm3.5\,$M$_{\rm Jup}$ and the outer orbital period to be $\sim$23.8$\,$yr, implying that the tertiary companion is either a massive brown dwarf or a very low-mass star in a wide orbit. We highlight that \RXGru{} has the highest outer-to-inner period ratio among the known PMS triples. The characteristics of \RXGru{} make it an ideal case study for exploring the early formation processes of short-period binaries in hierarchical triple systems. Thus, our results suggest that \RXGru{} was likely formed via the dynamical unfolding mechanism coupled with the shared accretion of the circumbinary material by the binary components.

%%%%%%%%%%%%%%%%%%%%%%%%%%%%%%%%%%%%%%%%%%%%%%%%%%

%%%%%%%%%%%%%%%%% ACKNOWLEDGEMENTS %%%%%%%%%%%%%%%

\begin{acknowledgements}

This work is based in part on data collected with Solaris network of telescopes of the Nicolaus Copernicus Astronomical Center of the Polish Academy of Sciences.
This work is based on observations collected at the European Organisation for Astronomical Research in the Southern Hemisphere under ESO programmes 60.A-9122 (eng.), 60.A-9036 (eng.), 077.D-0085, 087.C-0012, and 089.C-0415. 
Some of the observations reported in this paper were obtained with the Southern African Large Telescope (SALT) under programme 2021-2-MLT-006 (PI: A.\ Moharana). Polish participation in SALT is funded by the MEiN grant No.\ 2021/WK/01. 
This paper includes data collected by the \TESS{} mission, which are publicly available from the Mikulski Archive for Space Telescopes (MAST). Funding for the \TESS{} mission is provided by NASA’s Science Mission directorate. 
This paper makes use of data from the DR1 of the WASP data \citep{2010A&A...520L..10B} as provided by the WASP consortium, and computational resources supplied by the project `e-Infrastruktura CZ' (e-INFRA CZ LM2018140) supported by the Ministry of Education, Youth and Sports of the Czech Republic.
This work has made use of data from the European Space Agency (ESA) mission \Gaia{} (\url{https://www.cosmos.esa.int/gaia}), processed by the \Gaia{} Data Processing and Analysis Consortium (DPAC; \url{https://www.cosmos.esa.int/web/gaia/dpac/consortium}). Funding for the DPAC has been provided by national institutions, in particular the institutions participating in the \Gaia{} Multilateral Agreement. 
This research has made use of NASA's Astrophysics Data System Bibliographic Services, the SIMBAD database, operated at CDS, Strasbourg, France and the VizieR catalogue access tool, CDS, Strasbourg, France. The original description of the VizieR service was published in \citet{2000A&AS..143...23O}.

We acknowledge support provided by the Polish National Science Center through grants 2017/27/B/ST9/02727, 2021/41/N/ST9/02746, 2021/43/B/ST9/02972, 2023/49/B/ST9/01671, and 2024/53/N/ST9/03885. AM acknowledges support from the UK Science and Technology Facilities Council (STFC) under grant number ST/Y002563/1.

Finally, we wish to thank the anonymous referee for comments that helped us to improve this paper.
      
\end{acknowledgements}

%%%%%%%%%%%%%%%%%%%%%%%%%%%%%%%%%%%%%%%%%%%%%%%%%%

% Please note that we have included the references below in
% order to compile the document, but we ask you to:
   \bibliographystyle{aa} % style aa.bst
   \bibliography{fredm} % your references Yourfile.bib

%%%%%%%%%%%%%%%%%%%%%%%%%%%%%%%%%%%%%%%%%%%%%%%%%%

%%%%%%%%%%%%%%%%% APPENDICES %%%%%%%%%%%%%%%%%%%%%
% Appendices must be placed after   \end{thebibliography}
% They will be placed automatically on a new page.
%%%%%%%%%%%%%%%%%%%%%%%%%%%%%%%%%%%%%%%%%%%%%%%%%%%%%%%%%%%%%%%
\begin{appendix}
%%%%%%%%%%%%%%%%%%%%%%%%%%%%%%%%%%%%%%%%%%%%%%%%%%%%%%%%%%%%%%%
% In the PDF output, floats should be placed
% under their own appendix, not before the title, nor after the
% title of the next appendix.

% In short appendices, onecolumn floats (\figure*
% or \table*) will generate a blank page.
% To prevent this behaviour, a few examples are provided here. 

% In case you have a lot of floating objects for little text and the 
% LaTeX engine moves the floats away from their context, the command
% \FloatBarrier of the “placeins” package will empty the
% float buffer and place all stored floats in the continuity.

% If you still encounter problems with wide floats placement,
% just use the onecolumn environment throughout the appendices.
%%%%%%%%%%%%%%%%%%%%%%%%%%%%%%%%%%%%%%%%%%%%%%%%%%%%%%%%%%%%%%%

%____________________________________________________________
%       Wide floats at the start of an appendix: first method
%-------------------------------------------------------------
% To prevent a blank page after the start of an appendix:
% - Switch to one \onecolumn first
% - Declare the section title
% - Declare the onecolumn float with the parameter [h!]
% - Revert to \twocolumn at the end of the section
\onecolumn

%%%%%%%%%%%%%%%%%%%%%%%%%%%%%%%%%%%%%%%%%%%%%%%%%%

%%%%%%%%%%%%%%%%% TIMES OF MINIMA %%%%%%%%%%%%%%

\section{Times of minima}\label{app:ecl_times}

In Table~\ref{tab:ecl_times}, we provide the times of primary and secondary minima of \RXGru{} derived from the \TESS{} (sectors~1 and 28), \Solaris{}, and \SWASP{} light curves using the procedure described in Section~\ref{sec:ecl_times}.

\begin{longtable}{@{}lrrrrc@{}}
            \caption{Times of primary and secondary minima of \RXGru{}.}
            \label{tab:ecl_times}\\
            \hline\hline
            Time & \multicolumn{1}{c}{Cycle} & \multicolumn{1}{c}{1$\sigma$ error} & \multicolumn{1}{c}{$\Delta_{\rm obs}$} & \multicolumn{1}{c}{$O-C$} & Survey \\
            BJD$-243\,0000$ & \multicolumn{1}{c}{no.} & \multicolumn{1}{c}{(d)} & \multicolumn{1}{c}{(s)} & \multicolumn{1}{c}{(s)} & \\
            \hline
            \endfirsthead
            \caption{continued.}\\
            \hline\hline
            Time & \multicolumn{1}{c}{Cycle} & \multicolumn{1}{c}{1$\sigma$ error} & \multicolumn{1}{c}{$\Delta_{\rm obs}$} & \multicolumn{1}{c}{$O-C$} & Survey \\
            BJD$-243\,0000$ & \multicolumn{1}{c}{no.} & \multicolumn{1}{c}{(d)} & \multicolumn{1}{c}{(s)} & \multicolumn{1}{c}{(s)} & \\
            \hline
            \endhead
            \hline
            \endfoot
            23897.540\,143  &   $-$5959.0    &    0.000\,317   &  $-$345.8   &  $-$139.3  &  \SWASP{} \\
            23907.575\,871  &   $-$5945.5    &    0.000\,436   &   $-$57.9   &     166.9  &  \SWASP{} \\
            23910.547\,608  &   $-$5941.5    &    0.000\,238   &  $-$129.2   &      95.5  &  \SWASP{} \\
            23916.492\,458  &   $-$5933.5    &    0.000\,337   &  $-$152.8   &      71.5  &  \SWASP{} \\
            23917.606\,092  &   $-$5932.0    &    0.000\,151   &  $-$245.7   &   $-$40.3  &  \SWASP{} \\
            23920.578\,497  &   $-$5928.0    &    0.000\,144   &  $-$259.3   &   $-$54.0  &  \SWASP{} \\
            23936.556\,536  &   $-$5906.5    &    0.000\,276   &  $-$214.5   &       8.8  &  \SWASP{} \\
            23942.501\,700  &   $-$5898.5    &    0.000\,193   &  $-$210.9   &      11.9  &  \SWASP{} \\
            23943.616\,699  &   $-$5897.0    &    0.000\,200   &  $-$186.0   &      18.0  &  \SWASP{} \\
            23961.452\,724  &   $-$5873.0    &    0.000\,332   &  $-$129.5   &      73.4  &  \SWASP{} \\
            23964.424\,727  &   $-$5869.0    &    0.000\,227   &  $-$177.8   &      25.0  &  \SWASP{} \\
            23965.536\,808  &   $-$5867.5    &    0.000\,256   &  $-$405.0   &  $-$183.4  &  \SWASP{} \\
            23975.571\,483  &   $-$5854.0    &    0.000\,124   &  $-$208.1   &    $-$6.0  &  \SWASP{} \\
            23976.314\,224  &   $-$5853.0    &    0.000\,157   &  $-$242.6   &   $-$40.6  &  \SWASP{} \\
            23989.317\,535  &   $-$5835.5    &    0.000\,229   &  $-$384.9   &  $-$164.8  &  \SWASP{} \\
            23993.407\,156  &   $-$5830.0    &    0.000\,114   &  $-$182.0   &      19.0  &  \SWASP{} \\
            23999.353\,225  &   $-$5822.0    &    0.000\,382   &  $-$100.3   &     100.3  &  \SWASP{} \\
            24000.464\,302  &   $-$5820.5    &    0.000\,328   &  $-$414.3   &  $-$194.9  &  \SWASP{} \\
            24002.324\,869  &   $-$5818.0    &    0.000\,212   &  $-$179.6   &      20.9  &  \SWASP{} \\
            24003.436\,565  &   $-$5816.5    &    0.000\,400   &  $-$440.0   &  $-$220.8  &  \SWASP{} \\
            24028.335\,683  &   $-$5783.0    &    0.000\,189   &  $-$102.0   &      96.8  &  \SWASP{} \\
            24032.420\,583  &   $-$5777.5    &    0.000\,256   &  $-$306.9   &   $-$89.6  &  \SWASP{} \\
            24035.393\,821  &   $-$5773.5    &    0.000\,211   &  $-$248.5   &   $-$31.4  &  \SWASP{} \\
            24038.366\,653  &   $-$5769.5    &    0.000\,353   &  $-$225.2   &    $-$8.2  &  \SWASP{} \\
            24270.598\,945  &   $-$5457.0    &    0.000\,228   &  $-$147.0   &      32.7  &  \SWASP{} \\
            24273.571\,139  &   $-$5453.0    &    0.000\,257   &  $-$178.8   &       0.6  &  \SWASP{} \\
            24292.522\,208  &   $-$5427.5    &    0.000\,121   &   $-$93.5   &     103.1  &  \SWASP{} \\
            24295.494\,551  &   $-$5423.5    &    0.000\,222   &  $-$112.4   &      83.9  &  \SWASP{} \\
            24296.607\,366  &   $-$5422.0    &    0.000\,251   &  $-$276.2   &   $-$98.9  &  \SWASP{} \\
            24298.467\,080  &   $-$5419.5    &    0.000\,135   &  $-$115.2   &      80.8  &  \SWASP{} \\
            24299.579\,930  &   $-$5418.0    &    0.000\,156   &  $-$276.0   &   $-$98.9  &  \SWASP{} \\
            24330.422\,636  &   $-$5376.5    &    0.000\,167   &   $-$70.6   &     122.4  &  \SWASP{} \\
            24333.395\,098  &   $-$5372.5    &    0.000\,289   &   $-$79.2   &     113.5  &  \SWASP{} \\
            24350.486\,805  &   $-$5349.5    &    0.000\,174   &  $-$124.4   &      66.6  &  \SWASP{} \\
            24353.459\,468  &   $-$5345.5    &    0.000\,145   &  $-$115.7   &      75.1  &  \SWASP{} \\
            24360.518\,261  &   $-$5336.0    &    0.000\,124   &  $-$205.6   &   $-$34.4  &  \SWASP{} \\
            24362.376\,723  &   $-$5333.5    &    0.000\,134   &  $-$152.8   &      37.0  &  \SWASP{} \\
            24365.349\,158  &   $-$5329.5    &    0.000\,244   &  $-$163.7   &      25.8  &  \SWASP{} \\
            24372.409\,044  &   $-$5320.0    &    0.000\,197   &  $-$159.3   &      10.7  &  \SWASP{} \\
            24378.354\,707  &   $-$5312.0    &    0.000\,283   &  $-$112.7   &      56.7  &  \SWASP{} \\
            24379.468\,517  &   $-$5310.5    &    0.000\,126   &  $-$190.5   &    $-$2.3  &  \SWASP{} \\
            27244.277\,879  &   $-$1455.5    &    0.000\,194   &      65.1   &  $-$118.2  &  \Solaris{} \\
            27244.650\,352  &   $-$1455.0    &    0.000\,081   &     143.1   &   $-$59.1  &  \Solaris{} \\
            27246.136\,938  &   $-$1453.0    &    0.000\,065   &     169.4   &   $-$32.8  &  \Solaris{} \\
            27271.404\,240  &   $-$1419.0    &    0.000\,039   &     214.9   &      11.6  &  \Solaris{} \\
            27272.147\,459  &   $-$1418.0    &    0.000\,105   &     221.8   &      18.3  &  \Solaris{} \\
            27274.376\,734  &   $-$1415.0    &    0.000\,041   &     209.1   &       5.6  &  \Solaris{} \\
            27284.037\,838  &   $-$1402.0    &    0.000\,089   &     233.2   &      29.3  &  \Solaris{} \\
            27306.331\,582  &   $-$1372.0    &    0.000\,042   &     192.6   &   $-$12.2  &  \Solaris{} \\
            27318.964\,992  &   $-$1355.0    &    0.000\,074   &     194.5   &   $-$10.8  &  \Solaris{} \\
            27329.368\,715  &   $-$1341.0    &    0.000\,041   &     173.5   &   $-$32.2  &  \Solaris{} \\
            27628.110\,434  &    $-$939.0    &    0.000\,074   &     109.7   &  $-$102.1  &  \Solaris{} \\
            27629.226\,571  &    $-$937.5    &    0.000\,051   &     232.9   &      39.9  &  \Solaris{} \\
            27667.126\,086  &    $-$886.5    &    0.000\,055   &     177.0   &   $-$16.0  &  \Solaris{} \\
            27673.070\,990  &    $-$878.5    &    0.000\,079   &     158.0   &   $-$35.0  &  \Solaris{} \\
            27694.992\,982  &    $-$849.0    &    0.000\,074   &     101.8   &  $-$109.9  &  \Solaris{} \\
            27968.469\,055  &    $-$481.0    &    0.000\,046   &     135.5   &   $-$70.5  &  \Solaris{} \\
            28000.424\,306  &    $-$438.0    &    0.000\,045   &     153.9   &   $-$51.0  &  \Solaris{} \\
            28325.548\,419  &      $-$0.5    &    0.000\,024   &     168.5   &       0.9  &  \TESS{} S01 \\
            28325.920\,384  &         0.0    &    0.000\,033   &     202.6   &      16.2  &  \TESS{} S01 \\
            28326.291\,508  &         0.5    &    0.000\,024   &     164.0   &    $-$3.5  &  \TESS{} S01 \\
            28326.663\,476  &         1.0    &    0.000\,042   &     198.4   &      12.1  &  \TESS{} S01 \\
            28327.034\,741  &         1.5    &    0.000\,036   &     172.0   &       4.6  &  \TESS{} S01 \\
            28327.406\,663  &         2.0    &    0.000\,046   &     202.4   &      16.1  &  \TESS{} S01 \\
            28327.777\,887  &         2.5    &    0.000\,036   &     172.6   &       5.1  &  \TESS{} S01 \\
            28328.149\,900  &         3.0    &    0.000\,049   &     210.7   &      24.5  &  \TESS{} S01 \\
            28328.521\,007  &         3.5    &    0.000\,030   &     170.7   &       3.4  &  \TESS{} S01 \\
            28328.892\,932  &         4.0    &    0.000\,063   &     201.4   &      15.3  &  \TESS{} S01 \\
            28329.264\,288  &         4.5    &    0.000\,041   &     182.9   &      15.5  &  \TESS{} S01 \\
            28329.636\,095  &         5.0    &    0.000\,032   &     203.3   &      17.2  &  \TESS{} S01 \\
            28330.007\,374  &         5.5    &    0.000\,038   &     178.2   &      10.9  &  \TESS{} S01 \\
            28330.750\,372  &         6.5    &    0.000\,040   &     165.9   &    $-$1.3  &  \TESS{} S01 \\
            28331.122\,252  &         7.0    &    0.000\,033   &     192.6   &       6.6  &  \TESS{} S01 \\
            28331.493\,660  &         7.5    &    0.000\,030   &     178.7   &      11.5  &  \TESS{} S01 \\
            28331.865\,351  &         8.0    &    0.000\,030   &     189.1   &       3.2  &  \TESS{} S01 \\
            28332.236\,832  &         8.5    &    0.000\,046   &     181.4   &      14.3  &  \TESS{} S01 \\
            28332.608\,743  &         9.0    &    0.000\,073   &     210.8   &      24.9  &  \TESS{} S01 \\
            28332.979\,872  &         9.5    &    0.000\,025   &     172.7   &       5.6  &  \TESS{} S01 \\
            28333.351\,754  &        10.0    &    0.000\,029   &     199.6   &      13.8  &  \TESS{} S01 \\
            28333.723\,168  &        10.5    &    0.000\,023   &     186.1   &      19.1  &  \TESS{} S01 \\
            28334.094\,802  &        11.0    &    0.000\,032   &     191.7   &       5.9  &  \TESS{} S01 \\
            28334.466\,144  &        11.5    &    0.000\,030   &     171.9   &       5.0  &  \TESS{} S01 \\
            28334.838\,014  &        12.0    &    0.000\,028   &     197.8   &      12.1  &  \TESS{} S01 \\
            28335.209\,289  &        12.5    &    0.000\,028   &     172.3   &       5.4  &  \TESS{} S01 \\
            28335.581\,148  &        13.0    &    0.000\,035   &     197.3   &      11.6  &  \TESS{} S01 \\
            28336.324\,137  &        14.0    &    0.000\,034   &     184.1   &    $-$1.5  &  \TESS{} S01 \\
            28336.695\,591  &        14.5    &    0.000\,026   &     174.1   &       7.4  &  \TESS{} S01 \\
            28337.067\,339  &        15.0    &    0.000\,023   &     189.5   &       4.0  &  \TESS{} S01 \\
            28337.438\,684  &        15.5    &    0.000\,043   &     170.0   &       3.3  &  \TESS{} S01 \\
            28337.810\,369  &        16.0    &    0.000\,051   &     180.0   &    $-$5.5  &  \TESS{} S01 \\
            28338.181\,795  &        16.5    &    0.000\,028   &     167.4   &       0.8  &  \TESS{} S01 \\
            28340.039\,825  &        19.0    &    0.000\,036   &     182.9   &    $-$2.4  &  \TESS{} S01 \\
            28340.411\,282  &        19.5    &    0.000\,025   &     173.2   &       6.7  &  \TESS{} S01 \\
            28340.783\,034  &        20.0    &    0.000\,069   &     188.9   &       3.6  &  \TESS{} S01 \\
            28341.154\,462  &        20.5    &    0.000\,030   &     176.6   &      10.2  &  \TESS{} S01 \\
            28341.526\,105  &        21.0    &    0.000\,034   &     182.9   &    $-$2.3  &  \TESS{} S01 \\
            28341.897\,551  &        21.5    &    0.000\,027   &     172.1   &       5.8  &  \TESS{} S01 \\
            28342.269\,300  &        22.0    &    0.000\,029   &     187.6   &       2.4  &  \TESS{} S01 \\
            28342.640\,640  &        22.5    &    0.000\,033   &     167.7   &       1.4  &  \TESS{} S01 \\
            28343.012\,393  &        23.0    &    0.000\,028   &     183.5   &    $-$1.6  &  \TESS{} S01 \\
            28343.383\,847  &        23.5    &    0.000\,020   &     173.5   &       7.2  &  \TESS{} S01 \\
            28343.755\,732  &        24.0    &    0.000\,047   &     200.7   &      15.6  &  \TESS{} S01 \\
            28344.127\,031  &        24.5    &    0.000\,063   &     177.2   &      11.0  &  \TESS{} S01 \\
            28344.498\,774  &        25.0    &    0.000\,034   &     192.1   &       7.1  &  \TESS{} S01 \\
            28344.870\,111  &        25.5    &    0.000\,030   &     171.9   &       5.7  &  \TESS{} S01 \\
            28345.241\,882  &        26.0    &    0.000\,031   &     189.4   &       4.4  &  \TESS{} S01 \\
            28345.613\,144  &        26.5    &    0.000\,036   &     162.7   &    $-$3.4  &  \TESS{} S01 \\
            28345.985\,007  &        27.0    &    0.000\,039   &     188.0   &       3.1  &  \TESS{} S01 \\
            28346.356\,253  &        27.5    &    0.000\,025   &     160.0   &    $-$6.1  &  \TESS{} S01 \\
            28346.728\,198  &        28.0    &    0.000\,042   &     192.4   &       7.5  &  \TESS{} S01 \\
            28347.099\,531  &        28.5    &    0.000\,042   &     171.9   &       5.9  &  \TESS{} S01 \\
            28347.471\,391  &        29.0    &    0.000\,044   &     196.9   &      12.1  &  \TESS{} S01 \\
            28347.842\,536  &        29.5    &    0.000\,049   &     160.2   &    $-$5.7  &  \TESS{} S01 \\
            28348.957\,690  &        31.0    &    0.000\,101   &     198.5   &      13.8  &  \TESS{} S01 \\
            28349.700\,767  &        32.0    &    0.000\,044   &     193.0   &       8.4  &  \TESS{} S01 \\
            28350.071\,738  &        32.5    &    0.000\,027   &     141.2   &   $-$24.5  &  \TESS{} S01 \\
            28350.443\,807  &        33.0    &    0.000\,026   &     184.4   &    $-$0.2  &  \TESS{} S01 \\
            28350.814\,922  &        33.5    &    0.000\,027   &     145.0   &   $-$20.7  &  \TESS{} S01 \\
            28351.187\,134  &        34.0    &    0.000\,036   &     200.5   &      16.0  &  \TESS{} S01 \\
            28351.558\,143  &        34.5    &    0.000\,039   &     151.9   &   $-$13.7  &  \TESS{} S01 \\
            28351.930\,229  &        35.0    &    0.000\,038   &     196.5   &      12.1  &  \TESS{} S01 \\
            28352.301\,187  &        35.5    &    0.000\,025   &     143.6   &   $-$22.0  &  \TESS{} S01 \\
            28352.673\,287  &        36.0    &    0.000\,031   &     189.4   &       5.0  &  \TESS{} S01 \\
            28353.044\,317  &        36.5    &    0.000\,028   &     142.7   &   $-$22.9  &  \TESS{} S01 \\
            29061.999\,325  &       990.5    &    0.000\,088   &      58.6   &   $-$32.6  &  \TESS{} S28 \\
            29062.371\,720  &       991.0    &    0.000\,129   &     129.8   &      19.9  &  \TESS{} S28 \\
            29062.742\,481  &       991.5    &    0.000\,095   &      59.9   &   $-$31.1  &  \TESS{} S28 \\
            29063.114\,929  &       992.0    &    0.000\,134   &     135.8   &      25.9  &  \TESS{} S28 \\
            29063.485\,641  &       992.5    &    0.000\,084   &      61.7   &   $-$29.3  &  \TESS{} S28 \\
            29063.857\,983  &       993.0    &    0.000\,122   &     128.2   &      18.5  &  \TESS{} S28 \\
            29064.228\,775  &       993.5    &    0.000\,107   &      61.1   &   $-$29.8  &  \TESS{} S28 \\
            29064.601\,089  &       994.0    &    0.000\,122   &     125.4   &      15.7  &  \TESS{} S28 \\
            29064.971\,947  &       994.5    &    0.000\,107   &      63.8   &   $-$27.0  &  \TESS{} S28 \\
            29065.344\,215  &       995.0    &    0.000\,137   &     124.1   &      14.5  &  \TESS{} S28 \\
            29065.714\,989  &       995.5    &    0.000\,122   &      55.3   &   $-$35.4  &  \TESS{} S28 \\
            29066.087\,227  &       996.0    &    0.000\,145   &     113.0   &       3.5  &  \TESS{} S28 \\
            29066.458\,206  &       996.5    &    0.000\,114   &      61.9   &   $-$28.7  &  \TESS{} S28 \\
            29066.830\,307  &       997.0    &    0.000\,160   &     107.8   &    $-$1.6  &  \TESS{} S28 \\
            29067.201\,309  &       997.5    &    0.000\,130   &      58.7   &   $-$31.8  &  \TESS{} S28 \\
            29067.573\,395  &       998.0    &    0.000\,114   &     103.2   &    $-$6.1  &  \TESS{} S28 \\
            29067.944\,435  &       998.5    &    0.000\,107   &      57.4   &   $-$33.0  &  \TESS{} S28 \\
            29068.316\,543  &       999.0    &    0.000\,156   &     103.8   &    $-$5.3  &  \TESS{} S28 \\
            29068.687\,550  &       999.5    &    0.000\,126   &      55.2   &   $-$35.1  &  \TESS{} S28 \\
            29069.059\,532  &      1000.0    &    0.000\,130   &      90.8   &   $-$18.3  &  \TESS{} S28 \\
            29069.430\,618  &      1000.5    &    0.000\,137   &      49.0   &   $-$41.2  &  \TESS{} S28 \\
            29069.802\,681  &      1001.0    &    0.000\,122   &      91.5   &   $-$17.5  &  \TESS{} S28 \\
            29070.173\,737  &      1001.5    &    0.000\,153   &      47.1   &   $-$43.1  &  \TESS{} S28 \\
            29070.545\,815  &      1002.0    &    0.000\,122   &      90.9   &   $-$18.0  &  \TESS{} S28 \\
            29070.916\,847  &      1002.5    &    0.000\,168   &      44.5   &   $-$45.5  &  \TESS{} S28 \\
            29071.288\,933  &      1003.0    &    0.000\,145   &      89.2   &   $-$19.8  &  \TESS{} S28 \\
            29071.659\,973  &      1003.5    &    0.000\,168   &      43.2   &   $-$46.7  &  \TESS{} S28 \\
            29072.032\,089  &      1004.0    &    0.000\,114   &      90.4   &   $-$18.3  &  \TESS{} S28 \\
            29072.403\,030  &      1004.5    &    0.000\,175   &      36.0   &   $-$53.8  &  \TESS{} S28 \\
            29072.775\,307  &      1005.0    &    0.000\,145   &      97.1   &   $-$11.5  &  \TESS{} S28 \\
            29073.146\,202  &      1005.5    &    0.000\,183   &      38.7   &   $-$51.0  &  \TESS{} S28 \\
            29073.518\,548  &      1006.0    &    0.000\,172   &     105.8   &    $-$2.7  &  \TESS{} S28 \\
            29073.889\,318  &      1006.5    &    0.000\,212   &      36.6   &   $-$53.0  &  \TESS{} S28 \\
            29075.375\,710  &      1008.5    &    0.000\,168   &      46.2   &   $-$43.3  &  \TESS{} S28 \\
            29075.747\,734  &      1009.0    &    0.000\,107   &      85.4   &   $-$22.8  &  \TESS{} S28 \\
            29076.118\,874  &      1009.5    &    0.000\,160   &      48.2   &   $-$41.1  &  \TESS{} S28 \\
            29076.490\,898  &      1010.0    &    0.000\,153   &      87.5   &   $-$20.6  &  \TESS{} S28 \\
            29076.862\,076  &      1010.5    &    0.000\,175   &      53.6   &   $-$35.7  &  \TESS{} S28 \\
            29077.233\,955  &      1011.0    &    0.000\,092   &      80.3   &   $-$27.7  &  \TESS{} S28 \\
            29077.605\,270  &      1011.5    &    0.000\,175   &      58.2   &   $-$30.9  &  \TESS{} S28 \\
            29077.976\,936  &      1012.0    &    0.000\,076   &      66.5   &   $-$41.4  &  \TESS{} S28 \\
            29078.348\,412  &      1012.5    &    0.000\,175   &      58.3   &   $-$30.7  &  \TESS{} S28 \\
            29078.720\,261  &      1013.0    &    0.000\,175   &      82.4   &   $-$25.4  &  \TESS{} S28 \\
            29079.091\,560  &      1013.5    &    0.000\,160   &      59.0   &   $-$29.9  &  \TESS{} S28 \\
            29079.463\,295  &      1014.0    &    0.000\,084   &      73.2   &   $-$34.5  &  \TESS{} S28 \\
            29079.834\,663  &      1014.5    &    0.000\,175   &      55.8   &   $-$33.1  &  \TESS{} S28 \\
            29080.206\,482  &      1015.0    &    0.000\,099   &      77.3   &   $-$30.4  &  \TESS{} S28 \\
            29080.577\,774  &      1015.5    &    0.000\,175   &      53.2   &   $-$35.5  &  \TESS{} S28 \\
            29080.949\,729  &      1016.0    &    0.000\,139   &      86.5   &   $-$21.0  &  \TESS{} S28 \\
            29081.320\,803  &      1016.5    &    0.000\,190   &      43.6   &   $-$45.1  &  \TESS{} S28 \\
            29081.692\,825  &      1017.0    &    0.000\,137   &      82.7   &   $-$24.8  &  \TESS{} S28 \\
            29082.064\,049  &      1017.5    &    0.000\,168   &      52.7   &   $-$35.9  &  \TESS{} S28 \\
            29082.435\,852  &      1018.0    &    0.000\,076   &      72.8   &   $-$34.5  &  \TESS{} S28 \\
            29082.807\,198  &      1018.5    &    0.000\,160   &      53.4   &   $-$35.1  &  \TESS{} S28 \\
            29083.179\,077  &      1019.0    &    0.000\,084   &      80.1   &   $-$27.1  &  \TESS{} S28 \\
            29083.550\,323  &      1019.5    &    0.000\,160   &      52.2   &   $-$36.2  &  \TESS{} S28 \\
            29083.922\,165  &      1020.0    &    0.000\,084   &      75.6   &   $-$31.6  &  \TESS{} S28 \\
            29084.293\,465  &      1020.5    &    0.000\,160   &      52.2   &   $-$36.1  &  \TESS{} S28 \\
            29084.665\,436  &      1021.0    &    0.000\,092   &      86.9   &   $-$20.2  &  \TESS{} S28 \\
            29085.036\,621  &      1021.5    &    0.000\,153   &      53.6   &   $-$34.6  &  \TESS{} S28 \\
            29085.408\,424  &      1022.0    &    0.000\,076   &      73.7   &   $-$33.2  &  \TESS{} S28 \\
            29085.779\,755  &      1022.5    &    0.000\,175   &      53.0   &   $-$35.1  &  \TESS{} S28 \\
            29086.151\,604  &      1023.0    &    0.000\,084   &      77.1   &   $-$29.8  &  \TESS{} S28 \\
            29086.522\,916  &      1023.5    &    0.000\,123   &      54.8   &   $-$33.2  &  \TESS{} S28 \\
            29086.894\,699  &      1024.0    &    0.000\,092   &      73.2   &   $-$33.6  &  \TESS{} S28 \\
            29467.382\,090  &      1536.0    &    0.000\,080   &      28.7   &   $-$25.4  &  \Solaris{} \\
            29502.309\,686  &      1583.0    &    0.000\,057   &      28.3   &   $-$20.7  &  \Solaris{} \\
            29507.511\,420  &      1590.0    &    0.000\,159   &       6.7   &   $-$41.5  &  \Solaris{} \\
            29519.402\,099  &      1606.0    &    0.000\,111   &      44.1   &    $-$2.4  &  \Solaris{} \\
            29525.347\,192  &      1614.0    &    0.000\,041   &      41.4   &    $-$4.2  &  \Solaris{} \\
            29528.319\,934  &      1618.0    &    0.000\,033   &      57.0   &      11.8  &  \Solaris{} \\
\end{longtable}
\tablefoot{Half-integer cycle numbers refer to secondary eclipses. $O-C$ refers to the ETV residuals.}

%%%%%%%%%%%%%%%%%%%%%%%%%%%%%%%%%%%%%%%%%%%%%%%%%%

%%%%%%%%%%%%%%%%% RADIAL VELOCITIES %%%%%%%%%%%%%%

\section{Radial velocities}

In Table~\ref{tab:RV_obs}, we present our RV measurements used in this analysis. They refer to two components of the inner binary A = Aa+Ab. The first five (before BJD 245\,4000) are based on archival spectra, while the rest is from our campaigns. 

\begin{table*}[h!]
    \centering
    %\begin{minipage}{78mm}
        \caption{Individual RV measurements of \RXGru{} used in this work. All values are given in km\,s$^{-1}$. The last column tells which instrument was used.}
        \label{tab:RV_obs}
        {
        \renewcommand{\arraystretch}{1.0}
        \begin{tabular}{@{}lrrrrc@{}}
            \hline\hline
            BJD$-240\,0000$ & \multicolumn{1}{c}{$v_{\rm Aa}$} & \multicolumn{1}{c}{$\sigma_{\rm Aa}$} & \multicolumn{1}{c}{$v_{\rm Ab}$} & \multicolumn{1}{c}{$\sigma_{\rm Ab}$} & Sp. \\
            \hline
            53191.768\,258  &  $-$149.840 &  2.072  &  125.954    &  1.753  &  HARPS  \\
            53191.789\,440  &  $-$144.716 &  2.080  &  121.857    &  1.292  &  HARPS  \\
            53254.779\,149  &  $-$74.609  &  1.371  &  50.573     &  1.956  &  FEROS  \\
            53254.800\,618  &  $-$99.216  &  0.706  &  74.343     &  1.008  &  FEROS  \\
            53896.846\,796  &  $-$68.618  &  2.075  &  41.380     &  1.853  &  HARPS  \\
            55722.863\,039  &  $-$159.162 &  1.780  &  126.070    &  2.035  &  HARPS  \\
            55811.629\,029  &  113.953    &  1.760  &  $-$142.141 &  1.357  &  HARPS  \\
            55813.807\,145  &  77.669     &  1.197  &  $-$106.289 &  1.431  &  HARPS  \\
            56137.859\,122  &  111.600    &  1.762  &  $-$139.489 &  2.118  &  HARPS  \\
            56138.653\,497  &  126.157    &  2.321  &  $-$159.338 &  2.833  &  HARPS  \\
            56138.751\,092  &  100.404    &  1.424  &  $-$125.180 &  1.686  &  HARPS  \\
            56179.560\,984  &  123.277    &  1.404  &  $-$152.275 &  2.100  &  HARPS  \\
            59698.909\,504  &  35.385     &  1.249  &  $-$62.318  &  1.413  &  CHIRON \\
            59807.623\,885  &  104.941    &  1.449  &  $-$141.908 &  1.229  &  HRS    \\
            59835.544\,857  &  $-$92.902  &  1.418  &  64.319     &  0.890  &  HRS    \\
%            59861.482\,781  & $-$148.808  &  1.183  & 114.887	  &  1.385  &  HRS    \\
%            60108.889\,167	& $-$161.759  &  2.181  & 126.301     &  1.914  &  CHIRON \\
%            60121.915\,714	&  129.770    &  1.763	& $-$158.857  &  2.020	&  CHIRON \\
%            60123.822\,160	& $-$139.918  &  1.285  & 114.592     &  1.846  &  CHIRON \\
            \hline
        \end{tabular}
        }
    %\end{minipage}
\end{table*}

%%%%%%%%%%%%%%%%%%%%%%%%%%%%%%%%%%%%%%%%%%%%%%%%%%

%%%%%%%%%%%%%%%%%% SED photometry %%%%%%%%%%%%%%%%

\section{SED photometry}\label{app:sed}

Table~\ref{tab:sed_phot} provides the list of flux measurements used for the SED fitting.

\begin{table*}
\centering
    \caption{Photometry used for the SED fitting.}
    \label{tab:sed_phot}
    \begin{tabular}{@{}lccccc@{}}
        \hline\hline
        SED filter & Central wavelength & Filter width & Flux & Error & Model flux \\
        & $(\mathrm{\mathring{A}})$ & $(\mathrm{\mathring{A}})$ & $(\mathrm{erg\,\mathring{A}^{-1}\,s^{-1}\,cm^{-2}})$ & $(\mathrm{erg\,\mathring{A}^{-1}\,s^{-1}\,cm^{-2}})$ & $(\mathrm{erg\,\mathring{A}^{-1}\,s^{-1}\,cm^{-2}})$ \\
        \hline
        GALEX.FUV      & 1529.01 & 134.0   & $-$12.5 & 0.079  & $-$14.3650 \\
        GALEX.NUV      & 2311.96 & 308.0   & $-$10.6 & 0.0087 & $-$10.7510 \\
        Cousins.U      & 3500.2  & 319.0   & $-$9.46 & 0.0087 & $-$9.4171  \\
        SDSS.u         & 3519.02 & 277.0   & $-$9.47 & 0.0087 & $-$9.4466  \\
        Cousins.B      & 4359.98 & 464.0   & $-$9.08 & 0.0087 & $-$9.0790  \\
        Johnson.B      & 4442.03 & 445.0   & $-$9.09 & 0.043  & $-$9.0611  \\
        SDSS.g         & 4819.97 & 622.0   & $-$9    & 0.0087 & $-$8.9967  \\
        GAIA.GAIA3.Gbp & 5035.99 & 1079.0  & $-$9.01 & 0.0087 & $-$8.9829  \\
        Cousins.V      & 5469.67 & 421.0   & $-$8.96 & 0.0087 & $-$8.9224  \\
        Johnson.V      & 5537.05 & 409.0   & $-$8.97 & 0.0087 & $-$8.9235  \\
        GAIA.GAIA3.G   & 5822.34 & 2026.0  & $-$8.95 & 0.0087 & $-$8.9753  \\
        SDSS.r         & 6246.98 & 631.0   & $-$8.89 & 0.0087 & $-$8.8868  \\
        Cousins.R      & 6468.99 & 648.0   & $-$8.92 & 0.0087 & $-$8.8903  \\
        GAIA.GAIA3.Grp & 7620.55 & 1462.0  & $-$8.94 & 0.0087 & $-$8.9335  \\
        SDSS.i         & 7634.91 & 645.0   & $-$8.89 & 0.0087 & $-$8.8977  \\
        Cousins.I      & 7885.95 & 475.0   & $-$8.92 & 0.0087 & $-$8.9161  \\
        SDSS.z         & 9017.94 & 663.0   & $-$8.98 & 0.0087 & $-$8.9477  \\
        2MASS.Ks       & 21637.9 & 1250.0  & $-$9.5  & 0.0095 & $-$9.5357  \\
        WISE.W1        & 33500.1 & 3300.0  & $-$10   & 0.0087 & $-$10.0331 \\
        WISE.W2        & 46000.2 & 5200.0  & $-$10.4 & 0.0087 & $-$10.4367 \\
        WISE.W3        & 115598  & 27550.0 & $-$11.5 & 0.0087 & $-$11.6285 \\
    \hline
    \end{tabular}
\end{table*}

%\section{Spot phase diagram for 250d period}

%\begin{figure}
%    \centering
%    \includegraphics[width=\columnwidth]{figures/Phased_250SPOTtrend_.pdf}
%    \caption{Caption}
%    \label{fig:250dspot}
%\end{figure}

%\section{MCMC sampling of light curve model}
%\label{app:mcmc}

%\begin{figure*}
%    \includegraphics[width=\textwidth]{figures/corner.pdf}
%    \label{fig:corner}
%\end{figure*}

\end{appendix}

\end{document}